\documentclass[10pt,twocolumn,twoside]{IEEEtran}

\usepackage{amsmath,amsthm,amssymb,amsfonts}
\usepackage{bm}
\usepackage{commath}
\usepackage{mathtools}
\usepackage{centernot}

\usepackage{xcolor}
\usepackage{graphicx}

\definecolor{MyRed}{RGB}{232, 86, 82}
\definecolor{MyBlue}{RGB}{61, 83, 203}

\usepackage{tikz,pgfplots}

\usetikzlibrary{shapes, shadows.blur, automata, arrows, fit}

\tikzset{%
    >=latex,
	vertex/.style={circle,fill=black!25,minimum size=10pt,inner sep=0pt}, 
	edge/.style={draw,thick,->},    
	condensed/.style={rectangle,draw=green!50!black,thick},
}

\usepackage[prependcaption,colorinlistoftodos]{todonotes}

\usepackage{hyperref,doi,url}
\hypersetup{colorlinks=true, linkcolor=blue, breaklinks=true, urlcolor=blue}

\usepackage[font=small,labelformat=simple]{subcaption}
\renewcommand\thesubfigure{(\alph{subfigure})}

\usepackage{hhline}

\usepackage[linesnumbered,ruled,vlined]{algorithm2e}
\usepackage{algorithmic}

\usepackage{enumitem}

\usepackage{sgamevar}

\renewcommand{\gamestretch}{1.45}
\newcommand{\ts}[1]{\colorbox{red!30}{\makebox[7pt]{#1}}}   
\newtheorem{defi}{\textbf{Definition}}
\newtheorem{thom}{\textbf{Theorem}}

\newtheorem{rek}{\textbf{Remark}}

\newtheorem{lema}{\textbf{Lemma}}

\newtheorem{pbm}{\textbf{Problem}}
\newtheorem{cor}{\textbf{Corollary}}

\newtheorem{exmp}{\textbf{Example}}

\newcommand{\defiref}[1]{Definition \ref{#1}}
\newcommand{\thomref}[1]{Theorem~\ref{#1}}

\newcommand{\algref}[1]{Algorithm \ref{#1}}

\newcommand{\corref}[1]{Corollary \ref{#1}}
\newcommand{\lemaref}[1]{Lemma \ref{#1}}

\newcommand{\jointstrategyset}{\mathcal{S}^{\textup{J}}}
\newcommand{\jointstrategyedge}{\mathcal{E}^{\textup{J}}}
\newcommand{\jointstrategygraph}{\mathcal{G}^{\textup{J}}}
\newcommand{\singlestrategyset}{\mathcal{S}}
\newcommand{\jointstrategy}{s^{\textup{J}}}
\newcommand{\jointwrgraph}{\mathcal{G}_\textup{WR}^{\textup{J}}}
\newcommand{\jointwredge}{\mathcal{E}_{\textup{WR}}^{\textup{J}}}
\newcommand{\jointstgraph}{\mathcal{G}_\textup{ST}^{\textup{J}}}
\newcommand{\jointstedge}{\mathcal{E}_{\textup{ST}}^{\textup{J}}}
\newcommand{\singlewrgraph}{\mathcal{G}_\textup{N}}
\newcommand{\singlewredge}{\mathcal{E}_{\textup{N}}}
\newcommand{\singletgraph}{\mathcal{G}_\textup{B}}
\newcommand{\singletedge}{\mathcal{E}_{\textup{B}}}
\newcommand{\jointallSEwr}{\mathbb{Q}_\textup{WR}^{\textup{J}}}
\newcommand{\jointallSEst}{\mathbb{Q}_\textup{ST}^{\textup{J}}}
\newcommand{\singleallSEwr}{\mathbb{Q}_\textup{N}}

\newcommand{\singleallSEt}{\mathbb{Q}_\textup{B}}
\newcommand{\jointstadjmatrix}{\bm{A}_{\textup{ST}}^{\textup{J}}}
\newcommand{\singletadjmatrix}{\bm{A}_{\textup{B}}}
\newcommand{\adjmatrixbar}{\bar{\bm{A}}}
\newcommand{\jointcomponentwr}{\mathbb{Q}_{\textup{WR}}^{\textup{J}\#}}
\newcommand{\singlecomponentwr}{\mathbb{Q}_{\textup{N}}^{\#}}
\newcommand{\jointcomponentst}{\mathbb{Q}_{\textup{ST}}^{\textup{J}\#}}
\newcommand{\singlecomponentt}{\mathbb{Q}^{\#}_{\textup{B}}}

\begin{document}


\title{Evaluation and Learning in Two-Player Symmetric Games via Best and Better Responses}



\author{Rui~Yan,
        Weixian~Zhang,
        Ruiliang~Deng,
        Xiaoming~Duan,
        Zongying~Shi,~\IEEEmembership{Member,~IEEE,}
        and~Yisheng~Zhong%
\thanks{The work of R. Yan was funded by the ERC under the European Union’s Horizon 2020 research and innovation programme (\href{http://www.fun2model.org}{FUN2MODEL}, grant agreement No.~834115). The work of W. Zhang, R. Deng, Z. Shi and Y. Zhong was supported by the Science and Technology Innovation 2030---Key Project of ``New Generation Artificial Intelligence'' under Grant 2020AAA0108200 and Guoqiang Research Institute of Tsinghua University. }%
\thanks{R. Yan is with the Department of Computer Science, University of Oxford, Oxford, UK. E-mail: {\tt\small rui.yan@cs.ox.ac.uk}}
\thanks{W. Zhang, R. Deng, Z. Shi and Y. Zhong are with the Department of Automation, Tsinghua University, Beijing 100084, China. E-mails: {\tt\small \{wx-zhang17, drl20\}@mails.tsinghua.edu.cn}, {\tt\small \{szy,zys-dau\}@mail.tsinghua.edu.cn}.}%
\thanks{X. Duan is with the Department of Automation, Shanghai Jiao Tong University, Shanghai 200240, China. E-mail: {\tt\small xduan@sjtu.edu.cn}.}%
\thanks{R. Yan and W. Zhang contributed equally to this work.}
}

\maketitle

\IEEEpeerreviewmaketitle
\maketitle
\begin{abstract}
    Artificial intelligence and robotic competitions are accompanied by a class of game paradigms in which each player privately commits a strategy to a game system which simulates the game using the collected joint strategy and then returns payoffs to players.  This paper considers the strategy commitment for two-player symmetric games in which the players' strategy spaces are identical and their payoffs are symmetric. First, we introduce two digraph-based metrics at a meta-level for strategy evaluation in two-agent reinforcement learning, grounded on \emph{sink equilibrium}.  The metrics rank the strategies of a single player and determine the set of strategies which are preferred for the private commitment. Then, in order to find the preferred strategies under the metrics, we propose two variants of the classical learning algorithm self-play, called \emph{strictly best-response} and \emph{weakly better-response self-plays}. By modeling learning processes as walks over joint-strategy response digraphs, we prove that the learnt strategies by two variants are preferred under two metrics, respectively. The preferred strategies under both two metrics are identified
    and adjacency matrices induced by one metric and one variant are connected. Finally, simulations are provided to illustrate the results.

\end{abstract}

\begin{IEEEkeywords}
    Game theory, best and better responses, self-play, sink equilibrium, multi-agent reinforcement learning
\end{IEEEkeywords}

\section{Introduction}\label{part:introduction}
    \emph{Problem description and motivation:}  Multi-agent reinforcement learning (MARL) has achieved many successes in solving sequential decision-making problems in  control~\cite{li2020nonzero-sum,kober2013reinforcement} and games \cite{BY-GA-SY:21}. In a multi-player competition scenario, each player learns a strategy locally through MARL algorithms and privately commits the learnt strategy to a game system which simulates the game using the collected joint strategy and returns payoffs to players \cite{VC-TS:06,BP-WS-PT-SZ:19}. Which strategies should be committed and how to find them in the absence of the strategies to be committed by other players, are long-standing challenges in this paradigm and have attracted many attentions \cite{yan2021policy,balduzzi2018re-evaluating,lanctot2017unified,YD-XY-XC-JW-HZ:21,rowland2019multiagent,omidshafiei2019alpha-rank}.
    
    This paper introduces two metrics for strategies of a single player in two-player symmetric games in which the players' strategies are identical and their payoffs are symmetric (e.g., \textit{Go} \cite{silver2018general}, chess, poker \cite{brown2018superhuman} and video games \cite{jaderberg2019human-level}). The metrics rank the strategies and thus help the players to commit proper strategies. We also aim at developing learning algorithms with a finite memory to find the preferred strategies according to the proposed metrics. To this end, we formulate the underlying game system as a stochastic game and then analyze it as a meta-game which focuses on high-level interactions between players. We further propose two digraph-based metrics, and then design two new learning algorithms which are able to find the preferred strategies.
    


		\emph{Literature review:} Two-player symmetric games have been a central interest of the recent development of MARL \cite{silver2018general,brown2018superhuman,jaderberg2019human-level}, where the payoff for playing a particular strategy depends only on the other strategy employed, not on who is playing it. In solving such games, strategy evaluation and learning have been a long-standing challenge~\cite{lanctot2017unified,balduzzi2018re-evaluating,omidshafiei2019alpha-rank,YD-XY-XC-JW-HZ:21,omidshafiei2020navigating} due to multidimensional learning goals,  nonstationary environment, intransitive behaviors (e.g., Rock--Paper--Scissors) and scalability issues on the strategy space. Additionally, it is more common in artificial intelligence and robotic competitions that each player learns locally before playing against the opponent and then commits a learnt strategy to play the game \cite{yan2021policy,VC-TS:06}. This paradigm further complicates the evaluation and learning before the commitment, as the strategy that will be committed by the other player is unknown.
		
		For the strategy evaluation, the \emph{Elo rating system}~\cite{elo1978rating} and \emph{TrueSkill}~\cite{herbrich2007trueskill} have been widely used for evaluating the skills or abilities of artificial intelligence \cite{vinyals2019grandmaster,jaderberg2019human-level}. Unfortunately, these methods cannot deal with intransitive behaviors between strategies~\cite{rowland2019multiagent}. Focusing on intransitivity issues, many works have proposed response-graph based evaluation methods \cite{omidshafiei2019alpha-rank,yan2021policy,omidshafiei2020navigating,balduzzi2018re-evaluating,rowland2019multiagent,YD-XY-XC-JW-HZ:21}. For example, the work \cite{omidshafiei2019alpha-rank} proposes $\alpha$-Rank as a graph-based game-theoretical solution to multi-agent evaluation, which leverages perturbed better-response dynamics to rank strategies. More recent extensions of $\alpha$-Rank can be found in \cite{YD-XY-XC-JW-HZ:21,rowland2019multiagent}. In \cite{omidshafiei2020navigating}, the authors show that the response-graph based strategy evaluation enables the creation of a landscape of games, quantifying relationships between games of varying sizes and characteristics. In \cite{yan2021policy}, \emph{sink equilibria} \cite{goemans2005sink}, defined through the walks over response digraphs, are used to design cycle-based and memory-based strategy metrics. Though appealing, the above methods evaluate joint strategies and cannot be applied to our cases, because each player here does not know the other player's strategy and commits its own strategy independently, i.e., the evaluation of strategies of single players. The problem of designing proper multi-agent evaluation methods that address these issues, is still open.


     Game theory is a powerful tool in synthesizing strategies at a high-level for multiple interacting self-interested players. Most game-theoretical learning algorithms are based on \emph{best-response dynamics} \cite{roughgarden2010algorithmic}, in which each player plays a best response to the carefully designed stationary environment alternately which are equivalent to walks over the underlying best-response digraphs.  In the classical \emph{self-play} \cite{samuel1959studies}, copies of the learning player play each other in a best-response pattern. In the well-known \emph{fictitious play}~\cite{brown1951iterative}, each player optimally reacts to the empirical frequency of the opponents' previous plays. In \emph{adaptive play} \cite{HPY:04}, each player recalls a finite number of previous strategies used by the opponents and then plays a best response to the mixture of strategies sampled from the former set. In \emph{double oracle} \cite{HBM-GJG-AB:03}, each player plays a best response to the Nash equilibrium of its opponent over all previous learned joint strategies. A unified approach, called \emph{policy-space response oracles (PSRO)} \cite{lanctot2017unified}, generalizes fictitious play and double oracle. In the perturbed iterated best-response \cite{BY-GA-SY:21,yan2021policy,GA-SY:16}, the strategy selection of the player who needs to play a best response, is slightly perturbed such that, with a small probability, the player explores in the strategy space and adopts other strategies. 
    
    Game-theoretic approaches typically study how to converge to “good" strategies in the form of different equilibria, under the assumption that the players are “rational” in some sense. For instance, best-response dynamics is known for its convergence to pure-strategy Nash equilibrium in finite potential games~\cite{monderer1996potential}. Fictitious play and double oracle are guaranteed to converge to the set of Nash equilibria in two-player zero-sum games \cite{brown1951iterative,HBM-GJG-AB:03}. Perturbed strictly best-response dynamics is able to find the joint strategies with maximum underlying metrics for a class of multi-player nonzero-sum games \cite{yan2021policy}. In \cite{DC-TM:21}, distributed differential games are designed to compute the stable strategies/controllers for multi-agent systems. Rational imitation dynamics can guarantee finite time convergence to an imitation equilibrium profile for spatial public goods games \cite{AG-PR-MC:21}. Log-linear learning \cite{HB-JRM:17,JRM-JSS:12,KP-BC-PNB-MA-JRM:21}, a well-known distributed learning algorithm, guarantees the emergent behavior optimizes the system-level objective for multi-player nonzero-sum games. In \cite{CE-KP:20}, how to select a player to play a best response is investigated to avoid undesirable equilibria for anticoordination network games. Converging to Nash equilibria with disturbance rejection is considered for networked games in \cite{ARB-LP:20}. Proximal dynamics is proved to have global convergence to a network equilibrium for a class of multiagent network games in \cite{SG:18}. Designing learning algorithms which can converge to the preferred strategies  under a given evaluation method is an ongoing research direction \cite{lanctot2017unified,yan2021policy}.

	\emph{Contributions:} In this paper we focus on strategy evaluation and learning algorithms from the perspective of a single player who needs to commit a strategy independently to two-player symmetric games abstracted from stochastic games using high-level interactions,
	in the absence of the strategy committed by the other player. 
 The main contributions are as follows.
	\begin{enumerate}[label=(\roman*)]
    \item We first provide two metrics for the strategies of a single player, based on digraphs over the strategy space. Our metrics,
    differently from those of \cite{balduzzi2018re-evaluating,rowland2019multiagent,omidshafiei2019alpha-rank,yan2021policy,YD-XY-XC-JW-HZ:21} which evaluate joint strategies, rank the strategies of one player.
    
    \item Then, we propose two variants of self-play \cite{samuel1959studies}, \emph{strictly best-response} and \emph{weakly better-response self-plays}, and prove that they are able to learn the preferred strategies of the proposed metrics. Compared with the classical self-play, the first variant changes strategies with additional positive increment and thus simplifies the strategy jumps, and the second is more practical as determining better-response strategies is easier than best-response strategies given learning players. 
    
    \item Finally, we identify the strategies preferred by both two metrics 
    and through digraph product, connect the adjacency matrices induced by one metric and one variant.
\end{enumerate}
	
	
	
    
     
    \emph{Paper organisation:} In Section \ref{part:preliminaries}, we model the two-agent RL as a two-player stochastic game and reformulate it, at a high-level, as a symmetric normal-form game, and present the problems. In Section \ref{sec:evaluation-learning}, two metrics are introduced, and two learning algorithms adapted from self-play are proposed. In Section \ref{part:main-results}, we connect the proposed metrics and learning algorithms. We provide an example in Section \ref{part:outgrowths-of-the-main-results} and conclude the article in Section \ref{part:conclusion}.
    
    
    \noindent\textbf{Notations.}
    Let $\mathbb{R}$ and $\mathbb{N}_{>0}$ be the set of reals and positive integers, respectively. 
    For any finite set $S$, let $|S|$ be its cardinality, $\Delta_S$ the set of probability distributions over $S$, and $2^S$ the power set of $S$. 
    All vectors are column vectors. Let $\bm{1}$ denote the vector with all elements equal to 1, $\bm{e}_i$ the $i$-th standard basis vector, and $\bm{I}$ the identity matrix (their dimensions will be clear from the context).
    The Kronecker product of two matrices $\bm{A}$ and $\bm{B}$ is denoted by $\bm{A}\otimes\bm{B}$. If $\bm{A}$, $\bm{B}$, $\bm{C}$ and $\bm{D}$ are matrices with appropriate dimensions, then the mixed-product property of the Kronecker product says that $(\bm{A} \otimes \bm{B}) (\bm{C} \otimes \bm{D}) = (\bm{A} \bm{C}) \otimes (\bm{B} \bm{D})$. Partial important notations are provided in Table \ref{tab:notation} as an aid to understand the paper, which will be also explained in more details later.
    
    \begin{table}
    \caption{Notation Table}
    \label{tab:notation}
    \centering
    \begin{tabular}{ p{.15\linewidth}  p{.70\linewidth} } 
        \hline
        Symbol & Description \\
        \hline
    $s$, $s_i$, $s^i$ & Strategies of a single player\\
    $\jointstrategy$, $\jointstrategy_i$ & Joint strategy of two players\\
    $\singlestrategyset$, $\jointstrategyset$ & Set of strategies and joint strategies\\
    $\mathbb{B}(s)$ & Set of best responses to a strategy $s$\\
    $\singletgraph$, $\singlewrgraph$ & Best-response and non-dominated digraphs\\
    $\singleallSEt$, $\singleallSEwr$ & Set of sink equilibria (SEs) over $\singletgraph$ and $\singlewrgraph$\\
    $\singlecomponentt$, $\singlecomponentwr$ & Set of strategies in the elements of $\singleallSEt$ and $\singleallSEwr$\\
    $\jointstgraph$, $\jointwrgraph$ & Joint strategy strictly best-response and joint strategy weakly better-response digraphs\\
    $\jointallSEst$, $\jointallSEwr$ & Set of SEs over $\jointstgraph$ and $\jointwrgraph$\\
    $\jointcomponentst$, $\jointcomponentwr$ & Set of strategies in the elements of $\jointallSEst$ and $\jointallSEwr$\\
        \hline
    \end{tabular}
    \end{table}


%

\section{Problem Statement}\label{part:preliminaries}
\subsection{Two-Player Stochastic Games and Normal-Form Games}

Stochastic games have long been applied in MARL to model interactions among self-interested agents (players) in a shared environment. Consider a two-player stochastic game $G = (N, \mathcal{X}, d, \{\mathcal{A}^i\}_{i\in N}, P, \{R^i\}_{i\in N}, \{\beta^i\}_{i\in N})$, where $N=\{1, 2\}$ is a set of players, $\mathcal{X}$ is a finite set of states, $d \in \Delta_{\mathcal{X}}$ is an initial distribution over the states with $d(x_0)$ representing the probability of starting the game from a state $x_0 \in \mathcal{X}$, and $\mathcal{A}^i$ is a finite set of actions for player $i\in N$. The function $P:\mathcal{X}\times\mathcal{A}^1\times\mathcal{A}^2\to\Delta_{\mathcal{X}}$ determines the probability $P(x'\mid x,a^1,a^2)$ of transition from state $x$ to $x'$ under joint action $(a^1, a^2) \in \mathcal{A}^1 \times \mathcal{A}^2$. The immediate reward of player $i$ is given by $R^i:\mathcal{X}\times\mathcal{A}^1\times\mathcal{A}^2\to\mathbb{R}$. The scalar $\beta^i\in(0,1)$ is the discount factor for player $i$.


For player $i \in N$, a stationary deterministic strategy $s^i$ is a function from $\mathcal{X}$ to $\mathcal{A}^i$ \cite{GA-SY:16}, that is, $a^i=s^i(x)$. Denote the set of such strategies by $\mathcal{S}^i$, which is finite and satisfies $|\mathcal{S}^i|=|\mathcal{A}^i|^{|\mathcal{X}|}$. Let $\jointstrategyset=\mathcal{S}^1\times\mathcal{S}^2$ denote the joint strategy space,
and a joint strategy $\jointstrategy=(s^1,s^2)\in\jointstrategyset$ is also referred to as $\jointstrategy=(s^i,s^{-i})$ for any $i\in N$. 

Given a joint strategy $\jointstrategy \in\jointstrategyset$, the expected accumulated discounted payoff of player $i$ is computed by
\[
J^i(\jointstrategy)=\sum_{x \in \mathcal{X}}d(x) \mathbb{E}\big[\sum_{t=0}^\infty(\beta^i)^t R^i(x(t), a^1(t), a^2(t))\mid x_0=x \big].
\]
It is well-known that if each player has a set of finitely many strategies, then the related stochastic game has a normal-form game (NFG) representation~\cite{hansen2004dynamic}. From this, since $\mathcal{S}^i$ is finite for all $i \in N$, the NFG representation of the game $G$ considered here can be described by the triple $(N,\jointstrategyset,\{J^i\}_{i\in N})$, where the payoff of player $i$ under a joint strategy $\jointstrategy \in \jointstrategyset$ is $J^i(\jointstrategy)$ for $i \in N$. This NFG representation of a stochastic game is also called \emph{meta-game} in many literature \cite{omidshafiei2019alpha-rank,KT-JP-ML-JL-TG:18,yan2021policy}. A meta-game is a simplified model of complex interactions which focuses on meta-strategies (or styles of play) rather than atomic actions \cite{WW-RD-GT-JK:02}. For example, meta-strategies in poker may correspond to ``passive/aggressive'' or ``tight/loose'' strategies.

The two-player NFG is \emph{symmetric}~\cite{osborne2004introduction} if the strategy spaces of the players are identical and the payoffs of players are symmetric, i.e., $\mathcal{S}^1=\mathcal{S}^2=:\mathcal{S}$ (consequently $\jointstrategyset= \singlestrategyset \times \singlestrategyset$), and $J^1(s_1, s_2)=J^2(s_2, s_1)$ for all $s_1 \in\singlestrategyset$ and $s_2 \in\singlestrategyset$. 
By $\mathcal{S}^1=\mathcal{S}^2$, the action spaces are identical, i.e., $\mathcal{A}^1=\mathcal{A}^2=:\mathcal{A}$.
Note that in two-player symmetric NFGs (S-NFGs), 
the payoffs for playing a particular strategy depend only on the other strategy employed, not on who is playing it. Many real-world games, e.g., \textit{Go} \cite{silver2018general}, chess and poker \cite{brown2018superhuman}, are interesting examples of two-player S-NFGs. This paper will focus on two-player S-NFGs.

\subsection{Problem of Interest}

In many areas, e.g., artificial intelligence and robotics, the two-player S-NFG $G$ is played as follows \cite{brown2018superhuman,omidshafiei2020navigating}.
Each player first privately selects a strategy from the strategy set $\singlestrategyset$ and commits it to the game system (for example, game simulators). Then, the game system is fully driven
by the joint strategy which cannot be modified by the players once it starts. After many episodes, the game system returns an (expected) payoff to each player (averaged over the episodes). Each player tries to maximize its returned expected payoff by choosing a proper strategy from $\singlestrategyset$, given that the strategy committed by the opponent is unknown. Noting this, the aim of this article is, given $G$, to present a game-theoretical solution to the following problems.

\begin{pbm}[Strategy evaluation]
Design evaluation metrics over the strategies in $\singlestrategyset$ to help each player commit a proper strategy in $\singlestrategyset$ in the absence of the committed strategy by the opponent, so as to maximize its payoff.
\end{pbm}

\begin{pbm}[Strategy learning]
Given an evaluation metric, can we find, through learning, the (part of) strategies which are preferred under the evaluation metric? 
\end{pbm}






\section{Strategy Evaluation and Learning}\label{sec:evaluation-learning}

This section introduces two graph-based metrics to evaluate the strategies in $\singlestrategyset$. Then, two strategy learning algorithms are proposed, where the first one is closely related to the classical self-play \cite{samuel1959studies} and the second is less well studied but more useful in practice.

\subsection{Sink Equilibrium}

Before presenting the strategy evaluation and learning, we first introduce a game-theoretical concept called \emph{sink equilibrium} first proposed by Goemans \emph{et al.} \cite{goemans2005sink}.
Let $\mathcal{G}$ be a digraph with the node set being either $\singlestrategyset$ or $\jointstrategyset$, and an associated edge set which will be clear in the context.

\begin{defi}[Sink strongly connected component]
    A \emph{strongly connected component (SCC)} of a digraph $\mathcal{G}$ is a \emph{maximal} subgraph in which there is a path in each direction between each pair of nodes of the subgraph. A \emph{sink strongly connected component (SSCC)} is an SCC with no outgoing edges.
\end{defi}

\begin{defi}[Sink equilibrium, \cite{goemans2005sink}]\label{def:sink-equilibrium}
    A set $Q$ of nodes in the digraph $\mathcal{G}$ is a \emph{sink equilibrium (SE)} of an S-NFG $G$ over $\mathcal{G}$, if there exists an SSCC of $\mathcal{G}$ which is the induced-subgraph from $Q$.
\end{defi}

\subsection{Strategy Evaluation}

Before playing the game $G$, each player needs to commit a strategy in $\singlestrategyset$. From one player's point of view, the strategies in $\singlestrategyset$ must be properly evaluated and ranked before the commitment. Next, we introduce two digraph-based metrics. More precisely, a \emph{metric} is a function assigning a real number to each strategy, and a larger number means a higher preference.

Recently, evaluating the strategies for multi-agent systems through digraphs has been attracting much attention (see for example, \cite{rowland2019multiagent,omidshafiei2019alpha-rank,omidshafiei2020navigating,yan2021policy}). One advantage of this method is that it can model the incentive of a player changing its strategy to improve the payoff, when the strategies of the others are fixed. Most of the existing works \cite{rowland2019multiagent,omidshafiei2020navigating,yan2021policy} that use digraphs for strategy evaluation assign numbers to joint strategies in $\jointstrategyset$ instead of strategies in $\singlestrategyset$, and thus they are not applicable in our scenarios because each player here has to select one strategy from $\singlestrategyset$ rather than $\jointstrategyset$ and the strategy committed by the opponent is unknown. In addition, evaluating strategies in $\singlestrategyset$ is easier, because $\singlestrategyset$ has far less elements than $\jointstrategyset$. Strategy evaluation over $\singlestrategyset$ has been used in \cite{omidshafiei2019alpha-rank}, but with different digraphs and without detailed analysis. In this paper we will evaluate the strategies in $\singlestrategyset$ through two digraphs.

The set of best responses to a strategy $s \in \singlestrategyset$ is defined as
\[\mathbb{B}(s)=\big\{s_1 \in \singlestrategyset \mid J^1(s_1,s)=\max_{s_2 \in \singlestrategyset}J^1(s_2, s)\big\},
\]
For $s_1, s_2 \in \singlestrategyset$, if not specified, the first strategy in $J^1(s_1, s_2)$ or $J^2(s_1, s_2)$ is for player $1$ and the second is for player $2$.

\begin{defi}[Best-response digraph]\label{def:single-population-sst}
    A \emph{best-response digraph} $\singletgraph = (\singlestrategyset, \singletedge)$ of an S-NFG $G$ is a digraph where each node represents a strategy $s \in \singlestrategyset$ and an edge $e_{s_1 s_2}$ from $s_1 \in \singlestrategyset$ to $s_2 \in \singlestrategyset$ exists in $\singletedge$ if and only if $s_2 \in \mathbb{B}(s_1)$.
\end{defi}

\begin{defi}[Non-dominated digraph]\label{defi:ss-wbr} A \emph{non-dominated digraph} $\singlewrgraph = (\singlestrategyset,\singlewredge)$ of an S-NFG $G$ is a digraph where each node represents a strategy $s \in \singlestrategyset$ and an edge $e_{s_1s_2}$ from $s_1\in\singlestrategyset$ to $s_2\in\singlestrategyset$ exists in $\singlewredge$ if and only if there exists a strategy $s \in \singlestrategyset$ such that $J^1(s_2,s)\ge J^1(s_1,s)$.
\end{defi}

Let $\singleallSEt \subseteq 2^{\singlestrategyset}$ and $\singleallSEwr \subseteq 2^{\singlestrategyset}$ be the set of the SEs over $\singletgraph$ and $\singlewrgraph$, respectively. We denote by $\singlecomponentt \subseteq \singlestrategyset$ and $\singlecomponentwr \subseteq \singlestrategyset$ the set of strategies contained in the elements of $\singleallSEt$ and $\singleallSEwr$, respectively.

\begin{defi}[Best-dominating metric]
    A function $M: \singlestrategyset \to \mathbb{R}$ is a \emph{best-dominating  (BD) metric} if $M(s_1) > M(s_2)$ for all pairs $s_1 \in \singlecomponentt$ and $s_2 \in \singlestrategyset \setminus \singlecomponentt$. In other words, the strategies in $\singlecomponentt$ are preferred under the BD metric.  
\end{defi}

\begin{rek}
    Note that in the best-response digraph $\singletgraph$, there is a transition from one strategy to another in $\singlestrategyset$ if and only if adopting the latter results in the maximal payoff when the opponent is using the former. In other words, a strategy is better than another strategy if and only if the latter is a best response to the former. The BD metric says that the strategies in the SEs over $\singletgraph$ are preferred than the ones not.  
\end{rek}

\begin{defi}[Non-dominated metric]
    A function $M: \singlestrategyset \to \mathbb{R}$ is a \emph{non-dominated (ND) metric} if $M(s_1) > M(s_2)$ for all pairs $s_1 \in \singlecomponentwr$ and $s_2 \in \singlestrategyset \setminus \singlecomponentwr$. The strategies in $\singlecomponentwr$ are preferred under the ND metric.  
\end{defi}

\begin{rek}
    Regarding the non-dominated digraph $\singlewrgraph$, there is a transition from one strategy to another in $\singlestrategyset$ if and only if the latter is not dominated by the former, i.e., there exists at least one strategy in $\singlestrategyset$ against which adopting the latter gains a greater or equal payoff than adopting the former. The ND metric implies that, the strategies in the SEs over $\singlewrgraph$ are preferred than the ones not.  
\end{rek}



\subsection{Strategy Learning}

Based on the previous two metrics for strategy evaluation, we present two learning algorithms to find the strategies in $\singlecomponentt$ or $\singlecomponentwr$ which by definition, are assigned greater numbers than the ones outside correspondingly. That is, these strategies are preferred to commit under the proposed metrics. 

In general, it is computationally intractable to enumerate all nodes in $\singletgraph$ or $\singlewrgraph$ and then to determine the strategies in $\singlecomponentt$ or $\singlecomponentwr$, because if the S-NFG $G$ is induced from a stochastic game as considered here, the number of nodes in these graphs is $|\singlestrategyset|=|\mathcal{A}|^{|\mathcal{X}|}$. Noting this, we propose two game-theoretical learning algorithms by adapting the well-known self-play \cite{samuel1959studies}. We emphasize that different from the strategy evaluation, the opponent strategies are accessible during the local learning.

\begin{algorithm}
\caption{Strictly best-response self-play}\label{alg:sbt-sp}
\KwData{$N = \{ 1, 2 \}$, initial joint strategy $\jointstrategy_0=(s_0^1, s_0^2)$, maximum episode $\tau_{\textup{max}}$, memory length $L$}
\KwResult{a set of joint strategies}

$\tau\gets 0$

\Repeat{
$\tau = \tau_{\textup{max}}$ }
{
Choose a player $i$ from $N$ randomly




Learn a strategy $s^* \in \singlestrategyset$ for player $i$ such that $s^* \in \mathbb{B}(s^{-i}_{\tau})$ and $J^i(s^*, s^{-i}_{\tau}) > J^i(\jointstrategy_{\tau})$. If such an $s^*$ is not found within a given time, then $s^* \leftarrow s^i_{\tau}$


$s_{\tau + 1}^i \gets s^*$, $s_{\tau + 1}^{-i} \gets s_{\tau}^{-i}$

$\jointstrategy_{\tau + 1} \gets (s_{\tau + 1}^i, s_{\tau + 1}^{-i})$

$\tau\gets\tau + 1$


}
\textbf{return:} $\{\jointstrategy_{\tau + 1}\}_{ \tau = \tau_{\textup{max} -L+1}}^{\tau_{\textup{max}}}$
\end{algorithm}

\begin{algorithm}
\caption{Weakly better-response self-play}\label{alg:wbr-sp}
\KwData{$N = \{ 1, 2 \}$, initial joint strategy $\jointstrategy_0=(s_0^1, s_0^2)$, maximum episode $\tau_{\textup{max}}$, memory length $L$}
\KwResult{a set of joint strategies}

$\tau \leftarrow 0$

\Repeat{ $\tau = \tau_{\textup{max}}$
}
{
Choose a player $i$ from $N$ randomly

Learn a strategy $s^* \in \singlestrategyset$ for player $i$ such that $J^i(s^*, s^{-i}_{\tau}) \ge J^i(\jointstrategy_{\tau})$ 


$s_{\tau + 1}^i \gets s^*$, $s_{\tau + 1}^{-i} \gets s_{\tau}^{-i}$

$\jointstrategy_{\tau + 1} \gets (s_{\tau + 1}^i, s_{\tau + 1}^{-i})$

$\tau\gets\tau + 1$

}
\textbf{return:} $\{\jointstrategy_{\tau + 1}\}_{ \tau = \tau_{\textup{max} -L+1}}^{\tau_{\textup{max}}}$
\end{algorithm}

With respect to the conventional self-play, the strictly best-response variant described in \algref{alg:sbt-sp}, differs in that the chosen player $i \in N$ at the episode $\tau$  deviates from its current strategy $s_{\tau}^i$ to a learnt strategy $s^*$ if and only if $s^*$ is not only a best response to the opponent's current strategy $s_{\tau}^{-i}$ but also results in a strictly greater payoff (lines 4 and 5). The other player follows its previous strategy. We assume that only the $L$ latest learnt joint strategies are stored due to a finite memory constraint. The weakly better-response variant described in \algref{alg:wbr-sp} allows the chosen player $i$ to change its strategy from $s_{\tau}^i$ to $s^*$ if and only if adopting $s^*$ gains a weakly better payoff than adopting $s_{\tau}^i$. The second variant is more useful in practice than both conventional self-play and the first variant, because learning a best-response strategy is generally harder and more time-consuming than a better-response strategy, and it is also inefficient to verify that a strategy is a best response for reinforcement learning. In order to investigate the joint strategy behaviors in two variants, we introduce joint strategy response digraphs to model the dynamics of learning process, as in \cite{omidshafiei2020navigating,yan2021policy,GA-SY:16}.




\begin{defi}[Joint strategy strictly best-response digraph]\label{def:strictly-best}
		A \emph{joint strategy strictly best-response digraph} $\jointstgraph = (\jointstrategyset,\jointstedge)$ of an S-NFG $G$ is a digraph where each node is a joint strategy $\jointstrategy \in \jointstrategyset$ and an edge $e_{\jointstrategy_1 \jointstrategy_2}$ from $\jointstrategy_1 \in \jointstrategyset$ to $\jointstrategy_2 \in\jointstrategyset$ exists in $\jointstedge$ if and only if $\jointstrategy_1$ and $\jointstrategy_2$ differ in exactly one player's strategy (say $i\in N$), $s_2^i \in \mathbb{B}(s_1^{-i})$ and $J^i(\jointstrategy_2)>J^i(\jointstrategy_1)$.
\end{defi}

\begin{defi}[Joint strategy weakly better-response digraph]\label{def:weakly-better}
		A \emph{joint strategy weakly better-response graph} $\jointwrgraph =(\jointstrategyset,\jointwredge)$ of an S-NFG $G$ is a digraph where each node represents a joint strategy $\jointstrategy \in \jointstrategyset$ and an edge $e_{\jointstrategy_1 \jointstrategy_2}$ from $\jointstrategy_1 \in \jointstrategyset$ to $\jointstrategy_2 \in \jointstrategyset$ exists in $\jointwredge$ if and only if $\jointstrategy_1$ and $\jointstrategy_2$ differ in exactly one player's strategy (say $i\in N$), and $J^i(\jointstrategy_2)\ge J^i(\jointstrategy_1)$.
\end{defi}

The learning processes of strictly best-response and weakly better-response self-plays are the walks over $\jointstgraph$ and $\jointwrgraph$, respectively. This idea has been used for modeling multi-agent strategy learning behaviors in \cite{rowland2019multiagent,omidshafiei2019alpha-rank,omidshafiei2020navigating,yan2021policy}, because the induced digraphs can keep track of the way that the players explore in the strategy space. Let $\jointallSEst \subseteq 2^{\jointstrategyset}$ and $\jointallSEwr \subseteq 2^{\jointstrategyset}$ be the set of the SEs over $\jointstgraph$ and $\jointwrgraph$, respectively. Then we denote by $\jointcomponentst \subseteq \singlestrategyset$ and $\jointcomponentwr \subseteq \singlestrategyset$ the set of strategies contained in the elements of $\jointallSEst$ and $\jointallSEwr$, respectively.

\begin{rek}
    For a sufficiently large $\tau_{\textup{max}}$, the set of strategies contained in the elements of the learnt joint strategies $\{\jointstrategy_{\tau + 1}\}_{ \tau = \tau_{\textup{max} -L+1}}^{\tau_{\textup{max}}}$ through strictly best-response and weakly better-response self-plays is a subset of $\jointcomponentst$ and $\jointcomponentwr$, respectively.
\end{rek}

Since the initial joint strategy $\jointstrategy_0$ is randomly chosen, then any strategy in $\jointcomponentst$ or $\jointcomponentwr$ might be learnt after a long run. Noting this, we next discuss whether two variants of self-play proposed in this paper can find the preferred strategies under two metrics, through investigating the relationships between four sets of strategies $\singlecomponentt$, $\singlecomponentwr$, $\jointcomponentst$ and $\jointcomponentwr$.

\section{Bridges between Evaluation and Learning}\label{part:main-results}

\begin{figure}[tbp]
    \centering
    \footnotesize
    \begin{tikzpicture}[
        >=latex,
        algebraic relationship/.style={thick, dashed, MyRed},
        sink relationship/.style={thick, MyBlue},
        response graph/.style={draw, rounded corners, fill=white,  minimum width=3.5em, minimum height=1em}
    ]
		\matrix (relationships) [row sep=5.2em, column sep=5.8em] {
		    \node[] (evaluation_label) {\textbf{Evaluation}}; & \node[] (learning_label) {\textbf{Learning}}; \\[-4.5em]
			\node[response graph] (BD) {BD metric}; & \node[response graph] (Strictly) {Strictly best-response self-play}; \\
			\node[response graph] (ND) {ND metric}; & \node[response graph] (Weakly) {Weakly better-response self-play}; \\
		};
		\draw[<->, sink relationship] ([xshift=.3em]BD.south) -- node[right, sloped, anchor=south]{\footnotesize \thomref{thm:se-relationship-gwr0-gt0}} ([xshift=.3em]ND.north);
		\draw[<->, sink relationship] (BD) -- node[above]{\footnotesize \thomref{thm:se-relationship-gst-t0}} (Strictly);
		\draw[<->, sink relationship] (BD) -- node[below]{\footnotesize \thomref{lemma:algebraic-relationship-gst-gt0}} (Strictly);
		\draw[<->, sink relationship] (ND) -- node[above]{\footnotesize \thomref{thm:se-relationship-gwr-wr0}} (Weakly);
		\draw[<->, sink relationship] (ND) -- node[above, sloped, anchor=south]{\footnotesize \corref{corollary:se-relationship-gwr0-gst}} (Strictly);
		
	\end{tikzpicture}
    \caption{Main results on strategy evaluation and learning.}
    \label{fig:main-results}
\end{figure}
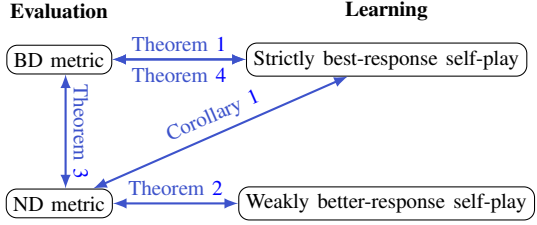


This section builds the bridges between the proposed evaluation metrics and learning algorithms, as depicted in Fig. \ref{fig:main-results}. First, we show that after many episodes, all learnt strategies by weakly better-response self-play are preferred under the ND metric, while all learnt strategies by strictly best-response self-play are preferred under the BD metric for a subclass of S-NFGs. Then, we compare the preferred strategies of the BD and ND metrics. Finally, the adjacency matrices related to the BD metric and strictly best-response self-play are connected.


\subsection{Learning Preferred Strategies}\label{part:se-ralationship}

In the following analysis, we assume that for two variants of self-play, the maximum episode $\tau_{\textup{max}}$ is large enough such that the walks over the joint strategy strictly best-response digraph $\jointstgraph$ or joint strategy weakly better-response digraph $\jointwrgraph$, have entered an SSCC and all $L$ latest learnt joint strategies belong to the corresponding SE over $\jointstgraph$ or $\jointwrgraph$.


We first connect the strategies preferred by the BD metric with the strategies returned by strictly best-response self-play.

\begin{figure}[t]
\centering
\footnotesize
\begin{subcaptiongroup}
	\begin{minipage}[b]{.45\linewidth}
		\hfil
		\sgcolsep=4pt\renewcommand{\gamestretch}{1.2}
		\begin{game}{3}{3}[\captiontext*{}]\phantomcaption\label{fig:btd-strictly-br-sp-self-payoff}
									\>	$s_1$	\>	$s_2$	\>	$s_3$	\\
									\vspace{10pt}
			$s_1$\hspace{\fboxsep}	\>	2, 2		\>	1, 2		\>	2, 1		\\
			$s_2$\hspace{\fboxsep}	\>	2, 1		\>	1, 1		\>	1, 2		\\
			$s_3$\hspace{\fboxsep}	\>	1, 2		\>	2, 2		\>	1, 1
		\end{game}\hspace{14.4pt} 
		\sgcolsep=1pt\renewcommand{\gamestretch}{1.45}
		\hfil\\
		\subcaptionbox{\label{fig:btd-strictly-br-sp-self-single-graph}}[\linewidth][c]{%
		\begin{tikzpicture}
			\node[vertex] (s1) at (-1.0,  0.0) {$s_1$};
			\node[vertex] (s2) at ( 1.0,  0.0) {$s_2$};
			\node[vertex] (s3) at ( 0.0, -0.7) {$s_3$};
			
			\draw[edge] (s1) to (s2);
			\draw[edge] (s2) to (s3);
			\draw[edge] (s3) to (s1);
			
			\draw[edge] (s1) to [loop left] (s1);
			\draw[edge,draw=none] (s2) to [loop right] (s2); 
		\end{tikzpicture}
		}
	\end{minipage}
	\begin{minipage}[b]{.45\linewidth}
		\subcaptionbox{\label{fig:btd-strictly-br-sp-self-joint-graph}}{%
			\begin{tikzpicture}
				\node[vertex] (s11) at (-1.0,  1.0) {$s_1s_1$};
				\node[vertex] (s12) at ( 0.0,  1.0) {$s_1s_2$};
				\node[vertex] (s13) at ( 1.0,  1.0) {$s_1s_3$};
				\node[vertex] (s21) at (-1.0,  0.0) {$s_2s_1$};
				\node[vertex] (s22) at ( 0.0,  0.0) {$s_2s_2$};
				\node[vertex] (s23) at ( 1.0,  0.0) {$s_2s_3$};
				\node[vertex] (s31) at (-1.0, -1.0) {$s_3s_1$};
				\node[vertex] (s32) at ( 0.0, -1.0) {$s_3s_2$};
				\node[vertex] (s33) at ( 1.0, -1.0) {$s_3s_3$};
				
				\draw[edge] (s12) to [bend right] (s32);
				
				\draw[edge] (s13) to [bend right=35] (s11);
				\draw[edge] (s13) to (s12);
				
				\draw[edge] (s21) to [bend left=35] (s23);
				
				\draw[edge] (s22) to (s23);
				\draw[edge] (s22) to (s32);
				
				\draw[edge] (s23) to (s13);
				
				\draw[edge] (s31) to [bend left=35] (s11);
				\draw[edge] (s31) to (s21);
				
				\draw[edge] (s32) to (s31);
				
				\draw[edge] (s33) to [bend left=45] (s13);
				\draw[edge] (s33) to [bend right=45] (s31);
			\end{tikzpicture}
		}
	\end{minipage}
\end{subcaptiongroup}
\caption{BD metric and strictly best-response self-play, where \subref{fig:btd-strictly-br-sp-self-payoff} payoffs of a two-player S-NFG; \subref{fig:btd-strictly-br-sp-self-single-graph} best-response digraph $\singletgraph$ with $\singleallSEt = \{ \{s_1, s_2, s_3\} \}$ and $\singlecomponentt = \{s_1, s_2, s_3\}$; \subref{fig:btd-strictly-br-sp-self-joint-graph} joint strategy strictly best-response digraph $\jointstgraph$ with $\jointallSEst=\{\{(s_1,s_1)\}\}$ and $\jointcomponentst = \{s_1\}$. Thus, $\jointcomponentst \subsetneq \singlecomponentt$.}\label{fig:SE-st-t0-2}
\end{figure}
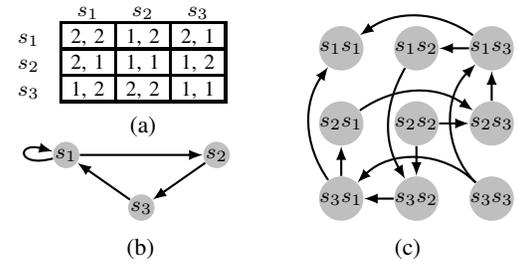

\begin{lema}[BD metric and strictly best-response self-play]\label{lema:se-relationship-gst-t0}
    For the BD metric and strictly best-response self-play,
    \begin{enumerate}[label=(\roman*)]
        \item \label{itm:st-t0-1} there exists a two-player S-NFG $G$ such that the strictly best-response self-play can return a strategy which is not preferred under the BD metric;
        
        \item \label{itm:st-t0-2} there exists a two-player S-NFG $G$ such that there exists a preferred strategy under the BD metric which cannot be returned by strictly best-response self-play.
    \end{enumerate}
\end{lema}
\begin{proof}
    Regarding \ref{itm:st-t0-1}, we consider a two-player S-NFG in Fig. \ref{fig:SE-st-t0-1}, where $\singlecomponentt = \{s_1, s_2\}$ and $\jointcomponentst = \{s_1, s_2, s_3\}$. Therefore, we have $\singlecomponentt \subsetneq \jointcomponentst$.
    
    Regarding \ref{itm:st-t0-2}, we consider a two-player S-NFG in Fig. \ref{fig:SE-st-t0-2}, where $\singlecomponentt = \{s_1, s_2, s_3\}$ and $\jointcomponentst = \{s_1\}$. Therefore, we have $\jointcomponentst \subsetneq \singlecomponentt$.
\end{proof}

\begin{figure}[t]
\centering
\footnotesize
\begin{subcaptiongroup}
	\begin{minipage}[b]{.45\linewidth}
		\hfil
		\sgcolsep=4pt\renewcommand{\gamestretch}{1.2}
		\begin{game}{3}{3}[\captiontext*{}]\phantomcaption\label{fig:btd-strictly-br-sp-mutual-payoff}
									\>	$s_1$   	\>	$s_2$	    \>	$s_3$       \\
			$s_1$\hspace{\fboxsep}	\>	2, 2		\>	1, 2		\>	1, 1		\\
			$s_2$\hspace{\fboxsep}	\>	2, 1		\>	1, 1		\>	2, 0		\\
			$s_3$\hspace{\fboxsep}	\>	1, 1		\>	0, 2		\>	2, 2
		\end{game}\hspace{14.4pt} 
		\sgcolsep=1pt\renewcommand{\gamestretch}{1.45}
		\hfil\\
		\subcaptionbox{\label{fig:btd-strictly-br-sp-mutual-single-graph}}[\linewidth][c]{%
		\begin{tikzpicture}
			\node[vertex] (s1) at (-1.0,  0.0) {$s_1$};
			\node[vertex] (s2) at ( 1.0,  0.0) {$s_2$};
			\node[vertex] (s3) at ( 0.0, -0.7) {$s_3$};
			
			\draw[edge] (s1) to [bend left=15] (s2);
			\draw[edge] (s2) to [bend left=15] (s1);
			\draw[edge] (s3) -- (s2);
			
			\draw[edge] (s1) to [loop left] (s1);
			\draw[edge] (s2) to [loop right] (s2);
			\draw[edge] (s3) to [loop left] (s3);
		\end{tikzpicture}
		}
	\end{minipage}
	\begin{minipage}[b]{.45\linewidth}
		\subcaptionbox{\label{fig:btd-strictly-br-sp-mutual-joint-graph}}{%
			\begin{tikzpicture}
				\node[vertex] (s11) at (-1.0,  1.0) {$s_1s_1$};
				\node[vertex] (s12) at ( 0.0,  1.0) {$s_1s_2$};
				\node[vertex] (s13) at ( 1.0,  1.0) {$s_1s_3$};
				\node[vertex] (s21) at (-1.0,  0.0) {$s_2s_1$};
				\node[vertex] (s22) at ( 0.0,  0.0) {$s_2s_2$};
				\node[vertex] (s23) at ( 1.0,  0.0) {$s_2s_3$};
				\node[vertex] (s31) at (-1.0, -1.0) {$s_3s_1$};
				\node[vertex] (s32) at ( 0.0, -1.0) {$s_3s_2$};
				\node[vertex] (s33) at ( 1.0, -1.0) {$s_3s_3$};
				
				\draw[edge] (s13) to [bend right=35] (s11);
				\draw[edge] (s13) to (s12);
				\draw[edge] (s13) to (s23);
				\draw[edge] (s13) to [bend right=35] (s33);
				
				\draw[edge] (s23) to [bend right=35] (s21);
				\draw[edge] (s23) to (s22);
				
				\draw[edge] (s31) to [bend left=35] (s11);
				\draw[edge] (s31) to (s21);
				\draw[edge] (s31) to (s32);
				\draw[edge] (s31) to [bend left=35] (s33);
				
				\draw[edge] (s32) to (s22);
				\draw[edge] (s32) to [bend left=35] (s12);
			\end{tikzpicture}
		}
	\end{minipage}
\end{subcaptiongroup}
\caption{BD metric and strictly best-response self-play, where \subref{fig:btd-strictly-br-sp-mutual-payoff} payoffs of a two-player S-NFG; \subref{fig:btd-strictly-br-sp-mutual-single-graph} best-response digraph $\singletgraph$ with $\singleallSEt = \{ \{s_1, s_2\} \} $ and $\singlecomponentt = \{s_1, s_2\}$; \subref{fig:btd-strictly-br-sp-mutual-joint-graph} joint strategy strictly best-response digraph $\jointstgraph$ with $\jointallSEst=\{\{(s_1,s_1)\},\{(s_1,s_2)\},\{(s_2,s_1)\},\{(s_2,s_2)\},\{(s_3,s_3)\}\}$, and $\jointcomponentst = \{s_1, s_2, s_3\}$. Thus, $\singlecomponentt \subsetneq \jointcomponentst$.}\label{fig:SE-st-t0-1}
\end{figure}

According to \lemaref{lema:se-relationship-gst-t0}, learning through the strictly best-response self-play may find the un-preferred strategies under the BD metric, thus providing bad candidates for the commitment. However, we are able to find that if a class of strategies are excluded (thus inducing a subclass of two-player S-NFGs), then $\jointcomponentst = \singlecomponentt$. Such strategies are defined as follows.


\begin{defi}[Self best-response strategy]
    A strategy $s \in \singlestrategyset$ is called \emph{a self best-response strategy} of a two-player S-NFG $G$ if $s\in\mathbb{B}(s)$, i.e., $(s,s)$ is a pure-strategy Nash equilibrium.
\end{defi}

\begin{defi}[Mutual best-response pair]
    A pair of strategy $\{s_1, s_2\}$ $(s_1, s_2 \in \singlestrategyset)$ is called \emph{a mutual best-response pair} of a two-player S-NFG $G$ if $s_1 \neq s_2$, $s_2\in\mathbb{B}(s_1)$ and $s_1\in\mathbb{B}(s_2)$.
\end{defi}

In order to prove the result, we require the following lemma.

\begin{lema}[Structure of $\singletgraph$]\label{lema:structure-t0}
If a two-player S-NFG $G$ has no self best-response strategies and no mutual best-response pairs, then
\begin{enumerate}[label=(\roman*)]
    \item\label{itm:three-node-t0} every SE over $\singletgraph$ has at least three distinct strategies;
    
    \item\label{itm:edge-t0} let $s_1$, $s_2$ and $s'$ be three distinct strategies in an SE over $\singletgraph$. If either $s_2 \in \mathbb{B}(s_1)$ or $s_1 \in \mathbb{B}(s_2)$, then there must exist a directed path in $\jointstgraph$ from $(s_1, s_2)$ to a joint strategy containing $s'$.
\end{enumerate}
\end{lema} 

\begin{proof}
    Regarding \ref{itm:three-node-t0}, if an SE (say $Q$) in $\singletgraph$ contains a unique strategy $s$, then by \defiref{def:single-population-sst} we have $s \in \mathbb{B}(s)$, which contradicts with no self best-response strategies. If $Q$ only contains two distinct strategies $s_1$ and $s_2$, there exist edges connecting them from both sides, that is, $\{s_1, s_2\}$ is a mutual best-response pair. Thus, the conclusion  \ref{itm:three-node-t0} holds.    
    
    Regarding \ref{itm:edge-t0}, we first focus on $s_2 \in \mathbb{B}(s_1)$ and the case $s_1 \in \mathbb{B}(s_2)$ follows similarly. Since $s_2$ and $s'$ are two distinct strategies in an SE (say $Q_1$) over $\singletgraph$, there must exist a best-response path from $s_2$ to $s'$. By attaching the best response from $s_1$ to $s_2$ at the beginning, this best-response path can be formulated as
    \begin{equation}\label{eq:strategy-1-prime}
        s_1 \rightarrow s_2 \rightarrow s_3 \rightarrow s_4 \rightarrow \cdots \rightarrow s_k \rightarrow s', 
    \end{equation}
    where $s_{i+1} \in \mathbb{B}(s_{i})$ for all $i = 1, 2 \dots, k -1$, and $s' \in \mathbb{B}(s_{k})$. Using \eqref{eq:strategy-1-prime}, we construct the following joint strategy path
    \begin{equation}\label{eq:strategy1-prime-1}
    (s_1, s_2) \rightarrow (s_3, s_2) \rightarrow (s_3, s_4) \rightarrow \cdots \rightarrow (s_k, s').
    \end{equation}
    We assume that the path \eqref{eq:strategy1-prime-1} ends up with $(s_k, s')$ and the case $(s', s_k)$ follows similarly. Since $G$ has no self best-response strategies, then any two strategies which are adjacent in the path \eqref{eq:strategy-1-prime} are distinct. Since $G$ has no mutual best-response pairs, then $s_1 \notin \mathbb{B}(s_2)$ (using $s_2 \in \mathbb{B}(s_1)$). Combining it with $s_3 \in \mathbb{B}(s_2)$, we have $J^1(s_3, s_2) > J^1 (s_1, s_2)$, implying that the edge from $(s_1, s_2)$ to $(s_3, s_2)$ is a strictly best response, i.e., $e_{(s_1, s_2)(s_3, s_2)} \in \jointstedge$. By the same argument until $(s_k, s')$, we are able to prove that \eqref{eq:strategy1-prime-1} is a joint strategy strictly best-response path, which as required, connects $(s_1, s_2)$ with a joint strategy containing $s'$. 
\end{proof}

We are ready to present the result connecting the strategies preferred by the BD metric with the ones returned by strictly best-response self-play for two-player S-NFGs which have no self best-response strategies and no mutual best-response pairs.

\begin{thom}[BD metric and strictly best-response self-play]\label{thm:se-relationship-gst-t0}
    If a two-player S-NFG $G$ has no self best-response strategies and no mutual best-response pairs, then the set of preferred strategies under the BD metric coincides with the set of all possible strategies returned by strictly best-response self-play, i.e., $\singlecomponentt = \jointcomponentst $.
\end{thom}


\begin{proof}
   Before proving the relation between $\singlecomponentt$ and $\jointcomponentst$, we first define a set of subgraphs of $\jointstgraph$ induced by the SEs in $\singletgraph$. For each SE $Q \in \singleallSEt$, let $\bar{\mathcal{G}}_{\textup{ST}}^{\textup{J}}(Q) = (\bar{{\jointstrategyset}}(Q), \bar{\mathcal{E}}_{\textup{ST}}^{\textup{J}}(Q))$ be the subgraph of $\jointstgraph$ such that $\bar{{\jointstrategyset}}(Q) = \{(s_1, s_2) \mid s_1 \in Q, s_2 \in Q \}$ and $e_{s_1^{\textup{J}}s_2^{\textup{J}}} \in \bar{\mathcal{E}}_{\textup{ST}}^{\textup{J}}(Q)$ if and only if $e_{s_1^{\textup{J}}s_2^{\textup{J}}} \in \jointstedge$ for all $s_1^{\textup{J}}, s_2^{\textup{J}} \in \bar{{\jointstrategyset}}(Q)$. Note that if $Q_1$ and $Q_2$ are two different SEs over $\singletgraph$, then $\bar{\mathcal{G}}_{\textup{ST}}^{\textup{J}}(Q_1)$ and $\bar{\mathcal{G}}_{\textup{ST}}^{\textup{J}}(Q_2)$ are two disjoint subgraphs of $\jointstgraph$, as $Q_1$ and $Q_2$ share no common strategies.
   
   For an SE $Q \in \singleallSEt$, we prove that any walk over $\jointstgraph$ cannot leave the subgraph $\bar{\mathcal{G}}_{\textup{ST}}^{\textup{J}}(Q)$ once entering it. We assume that there exists an edge in $\jointstedge$ leaving $\bar{\mathcal{G}}_{\textup{ST}}^{\textup{J}}(Q)$, i.e., there exist three strategies $s_1 \in \singlestrategyset \setminus Q$, $s_2 \in  Q$ and $s_3 \in Q$ ($s_2$ and $s_3$ might be the same) such that $s_1 \in \mathbb{B}(s_2)$ and $J^1(s_1, s_2) > J^1(s_3, s_2)$. However, in view of $s_1 \in \singlestrategyset \setminus Q$ and $s_2 \in  Q$, we obtain the contradiction $s_1 \notin \mathbb{B}(s_2)$ as $Q$ is an SE over $\singletgraph$. 
   
   Now we are ready to prove the relation. Suppose $s_1 \in \jointcomponentst$. We assume, for the sake of contradiction, that $s_1 \notin \singlecomponentt$, i.e., there exists a best-response path over $\singletgraph$ from $s_1$ to strategies in an SE (say $Q_1$) over $\singletgraph$. More precisely, this best-response path can be formulated as
   \begin{equation}\label{eq:path-s0-strategy-1}
    \begin{aligned}
         s_1 \rightarrow s_2 \rightarrow s_3 \rightarrow s_4 \rightarrow \cdots \rightarrow s_k \rightarrow s_{k+1},
    \end{aligned}
   \end{equation}
   where $s_{i+1} \in \mathbb{B}(s_{i})$ for all $i = 1, 2 \dots, k$, $s_1 \notin Q_1$, $s_k \in Q_1$ and $s_{k+1} \in Q_1$ by noting that $Q_1$ has at least three distinct strategies by \ref{itm:three-node-t0} in \lemaref{lema:structure-t0}. Since $G$ has no self best-response strategies, then $s_{i}$ and $s_{i+1}$ are different for all $i = 1, 2 \dots, k$. 
   
   Next, we prove that every joint strategy in $\jointstrategyset$ containing $s_1$ has a joint strategy strictly best-response path to a subgraph $\bar{\mathcal{G}}_{\textup{ST}}^{\textup{J}}(Q)$ induced by some SE $Q \in \singleallSEt$. Given a strategy $s' \in \singlestrategyset$, we consider the joint strategy $(s', s_1)$, and the case $(s_1, s')$ follows similarly. Using \eqref{eq:path-s0-strategy-1}, we consider the following joint strategy path
   \begin{equation}\label{eq:s-prime-joint-stra}
    \begin{aligned}
        (s', s_1) \rightarrow (s_2, & s_1) \rightarrow (s_2, s_3) \\
        & \rightarrow (s_4, s_3) \rightarrow \cdots \rightarrow (s_k, s_{k+1}).
    \end{aligned}
   \end{equation}
    If \eqref{eq:s-prime-joint-stra} ends up with $(s_{k+1}, s_k)$, we can also follow the same argument as below. There are three cases with respect to $s'$, and we next discuss them separately.
    
    Case 1: $s' \notin \mathbb{B}(s_1)$. Since $s_2 \in \mathbb{B}(s_1)$, we have $J^1(s_2, s_1) > J^1(s', s_1)$, implying that the edge from $(s', s_1)$ to $(s_2, s_1)$ is a joint strategy strictly best response, i.e., $e_{(s', s_1)(s_2, s_1)} \in \jointstedge$. Since $G$ has no mutual best-response pair, then $s_1 \notin \mathbb{B}(s_2)$ (using $s_2 \in \mathbb{B}(s_1)$). Then combining it with $s_3 \in \mathbb{B}(s_2)$, we have $J^2(s_2, s_3) > J^2 (s_2, s_1)$, implying that the edge from $(s_2, s_1)$ to $(s_2, s_3)$ is a joint strategy strictly best response, i.e., $e_{(s_2, s_1)(s_2, s_3)} \in \jointstedge$. By the same argument until $(s_k, s_{k+1})$, we are able to prove that \eqref{eq:s-prime-joint-stra} is a joint strategy strictly best-response path. Recall that $s_1 \notin Q_1$, $s_k \in Q_1$ and $s_{k+1} \in Q_1$. Therefore, all joint strategies in the subgraph $\bar{\mathcal{G}}_{\textup{ST}}^{\textup{J}}(Q_1)$ don't contain $s_1$, $(s_k, s_{k+1})$ is included in $\bar{\mathcal{G}}_{\textup{ST}}^{\textup{J}}(Q_1)$ and $(s', s_1)$ is at the outside of $\bar{\mathcal{G}}_{\textup{ST}}^{\textup{J}}(Q_1)$. Since any walk over $\jointstgraph$ cannot leave  $\bar{\mathcal{G}}_{\textup{ST}}^{\textup{J}}(Q_1)$ once entering it, then $(s', s_1)$ doesn't lie in any SE over $\jointstgraph$, noting that it can enter $\bar{\mathcal{G}}_{\textup{ST}}^{\textup{J}}(Q_1)$ along \eqref{eq:s-prime-joint-stra}.
    
    Case 2: $s' \in \mathbb{B}(s_1)$ and $s' = s_2$. Then, we consider the joint strategy path 
    \begin{equation*}
       (s_2, s_1) \rightarrow (s_2, s_3) \rightarrow (s_4, s_3) \rightarrow \cdots \rightarrow (s_k, s_{k+1}),
   \end{equation*}
   and prove that $(s_2, s_1)$ doesn't lie in any SE over $\jointstgraph$ similarly. 
   
   Case 3: $s' \in \mathbb{B}(s_1)$ and $s' \neq s_2$. Since $s_1 \notin \singlecomponentt$, then there exists a best-response path over $\singletgraph$ from $s_1$ to $s'$ and then from $s'$ to strategies in an SE in $\singleallSEt$. This implies that we have a path similar to \eqref{eq:path-s0-strategy-1} and thus can prove that $(s', s_1)$ doesn't lie in any SE over $\jointstgraph$ by following the similar argument to Case 2.
   
   Since $s'$ can be any strategy in $\singlestrategyset$, we obtain the contradiction $s_1 \notin \jointcomponentst$. Then, $s_1 \in \singlecomponentt$ and thus $ \jointcomponentst \subset \singlecomponentt$.
   
    Conversely, suppose $s_1 \in \singlecomponentt$, i.e., there exists an SE $Q_1 \in \singleallSEt$ such that $s_1 \in Q_1$. We assume, for the sake of contradiction, that $s_1 \notin \jointcomponentst$. This implies that there exists a joint strategy strictly best-response path over $\jointstgraph$ from $(s_1, s_1)$ to a joint strategy in an SE (say $Q^{\textup{J}}_1$) in $\jointallSEst$ and all joint strategies in this SE $Q^{\textup{J}}_1$ don't contain $s_1$. More precisely, this joint strategy strictly best-response path from $(s_1, s_1)$ to a joint strategy in $Q^{\textup{J}}_1$ can be formulated as follows
   \begin{equation}\label{eq:path-s0-joint-strategy}
    \begin{aligned}
        (s_1, s_1) \rightarrow (s_2, s_1) \rightarrow (s_2, s_3) \rightarrow (s_4, s_3) \rightarrow \cdots \rightarrow (s_k, s_{k+1}),
    \end{aligned}
   \end{equation}
   where $s_{i+1} \in \mathbb{B}(s_{i})$ for all $i = 1, 2 \dots, k$, and there is a strict payoff improvement for the player playing the best response at each edge and $(s_k, s_{k+1}) \in Q^{\textup{J}}_1$. We can construct the path \eqref{eq:path-s0-joint-strategy} with the following reasons: Firstly, the first edge is because $G$ has no self best-response strategies (i.e., $s_1$ is not a best response of itself); secondly, two players must play strictly best responses alternately along a path over $\jointstgraph$ using \defiref{def:strictly-best}; thirdly, all joint strategies in the path except the first one $(s_1, s_1)$ have different strategies for two players because of no self best-response strategies. If the path starts with player $2$ playing the best response first or ends up with $(s_{k+1}, s_{k})$, we can have the same argument as below. According to these features, \eqref{eq:path-s0-joint-strategy} induces a best-response path over $\singletgraph$:
   \begin{equation}
    \begin{aligned}
        s_1 \rightarrow s_2 \rightarrow s_3 \rightarrow s_4 \rightarrow \cdots \rightarrow s_k \rightarrow s_{k+1}.
    \end{aligned}
   \end{equation}
   Recall that $s_1 \in Q_1$. Then, we have $s_k \in Q_1$ and $s_{k+1} \in Q_1$. Since no joint strategy in $Q^{\textup{J}}_1$ contains $s_1$ and $(s_k, s_{k+1}) \in Q^{\textup{J}}_1$, then $s_1$ is distinct from $s_k$ and $s_{k+1}$. Since $s_{k}$ is different from $s_{k+1}$ and $s_{k+1} \in \mathbb{B}(s_k)$, then by \ref{itm:edge-t0} in \lemaref{lema:structure-t0}, there exists a joint strategy strictly best-response path in $\jointstgraph$ from $(s_k, s_{k+1})$ to a joint strategy (say $s^{\textup{J}}$) containing $s_1$, which implies that $s^{\textup{J}} \in Q^{\textup{J}}_1$. This contradicts with the fact that no joint strategy in $Q^{\textup{J}}_1$ contains $s_1$. Then, we have $s_1 \in \jointcomponentst$ and thus $\singlecomponentt \subset \jointcomponentst$.
\end{proof}

Next we connect the strategies preferred by the ND metric with the strategies returned by weakly better-response self-play for two-player S-NFGs. Before presenting the main results, the following lemma is required.

\begin{lema}[SE in $\singlewrgraph$]\label{lema:structure-wr0}
For a two-player S-NFG $G$, $\singlewrgraph$ admits a unique SE.
    
\end{lema}

\begin{proof}
    For any $s_1, s_2 \in \singlestrategyset$, we have $J^1(s_1, s_1) \leq J^1(s_2, s_1)$ or $J^1(s_1, s_1) \ge J^1(s_2, s_1)$, i.e., either $e_{s_1s_2} \in \singlewredge$ or $e_{s_2s_1} \in \singlewredge$. Therefore, there exists at least one edge between any two nodes in $\singlewrgraph$.
    
    Suppose that $\singlewrgraph$ has more than one SE. Let $Q_1$ and $Q_2$ be two distinct SEs in $\singlewrgraph$. Take one node from each SE, say $s_1 \in Q_1$ and $s_2 \in Q_2$. Since there exists at least one edge between $s_1$ and $s_2$, implying that $Q_1$ and $Q_2$ cannot be two distinct SEs. Therefore, $\singlewrgraph$ admits a unique SE (the existence is obvious).
\end{proof}

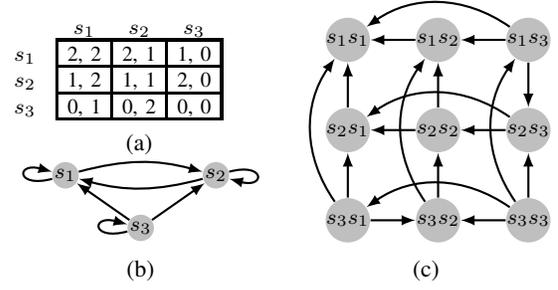
\begin{figure}[t]
\centering
\footnotesize
\begin{subcaptiongroup}
	\begin{minipage}[b]{.45\linewidth}
		\hfil
		\sgcolsep=4pt\renewcommand{\gamestretch}{1.2}
		\begin{game}{3}{3}[\captiontext{}]\phantomcaption\label{fig:brd-weakly-br-sp-self-payoff}
									\>	$s_1$	\>	$s_2$	\>	$s_3$	\\
			$s_1$\hspace{\fboxsep}	\>	2, 2		\>	2, 1		\>	1, 0		\\
			$s_2$\hspace{\fboxsep}	\>	1, 2		\>	1, 1		\>	2, 0		\\
			$s_3$\hspace{\fboxsep}	\>	0, 1		\>	0, 2		\>	0, 0
		\end{game}\hspace{14.4pt} 
		\sgcolsep=1pt\renewcommand{\gamestretch}{1.45}
		\hfil\\
		\subcaptionbox{\label{fig:brd-weakly-br-sp-self-single-graph}}[\linewidth][c]{%
		\begin{tikzpicture}
			\node[vertex] (s1) at (-1.0,  0.0) {$s_1$};
			\node[vertex] (s2) at ( 1.0,  0.0) {$s_2$};
			\node[vertex] (s3) at ( 0.0, -0.7) {$s_3$};
			
			\draw[edge] (s1) to [bend left=15] (s2);
			\draw[edge] (s2) to [bend left=15] (s1);
			\draw[edge] (s3) -- (s1);
			\draw[edge] (s3) -- (s2);
			
			\draw[edge] (s1) to [loop left] (s1);
			\draw[edge] (s2) to [loop right] (s2);
			\draw[edge] (s3) to [loop left] (s3);
		\end{tikzpicture}
		}
	\end{minipage}
	\begin{minipage}[b]{.45\linewidth}
		\subcaptionbox{\label{fig:brd-weakly-br-sp-self-joint-graph}}{%
			\begin{tikzpicture}
				\node[vertex] (s11) at (-1.2,  1.2) {$s_1s_1$};
				\node[vertex] (s12) at ( 0.0,  1.2) {$s_1s_2$};
				\node[vertex] (s13) at ( 1.2,  1.2) {$s_1s_3$};
				\node[vertex] (s21) at (-1.2,  0.0) {$s_2s_1$};
				\node[vertex] (s22) at ( 0.0,  0.0) {$s_2s_2$};
				\node[vertex] (s23) at ( 1.2,  0.0) {$s_2s_3$};
				\node[vertex] (s31) at (-1.2, -1.2) {$s_3s_1$};
				\node[vertex] (s32) at ( 0.0, -1.2) {$s_3s_2$};
				\node[vertex] (s33) at ( 1.2, -1.2) {$s_3s_3$};
				
				\draw[edge] (s12) to (s11);
				
				\draw[edge] (s13) to [bend right=35] (s11);
				\draw[edge] (s13) to (s12);
				\draw[edge] (s13) to (s23);
				
				\draw[edge] (s21) to (s11);
				
				\draw[edge] (s22) to (s12);
				\draw[edge] (s22) to (s21);
				
				\draw[edge] (s23) to [bend right=35] (s21);
				\draw[edge] (s23) to (s22);
				
				\draw[edge] (s31) to [bend left=35] (s11);
				\draw[edge] (s31) to (s21);
				\draw[edge] (s31) to (s32);
				
				\draw[edge] (s32) to [bend left=35] (s12);
				\draw[edge] (s32) to (s22);
				
				\draw[edge] (s33) to [bend left=35] (s13);
				\draw[edge] (s33) to (s23);
				\draw[edge] (s33) to [bend right=35] (s31);
				\draw[edge] (s33) to (s32);
			\end{tikzpicture}
		}
	\end{minipage}
\end{subcaptiongroup}
\caption{ND metric and weakly better-response self-play, where \subref{fig:brd-weakly-br-sp-self-payoff} payoffs of a two-player S-NFG; \subref{fig:brd-weakly-br-sp-self-single-graph} non-dominated digraph $\singlewrgraph$ with $\singleallSEwr = \{ \{ s_1, s_2 \} \}$ and $\singlecomponentwr = \{s_1, s_2\}$; \subref{fig:brd-weakly-br-sp-self-joint-graph} joint strategy weakly better-response digraph $\jointwrgraph$ with $\jointallSEwr=\{\{(s_1,s_1)\}\}$ and $\jointcomponentwr = \{s_1\}$. Thus, $\jointcomponentwr \subsetneq \singlecomponentwr$.}\label{fig:SE-wr-wr0}

\end{figure}

\begin{thom}[ND metric and weakly better-response self-play]\label{thm:se-relationship-gwr-wr0}
    Consider a two-player S-NFG $G$. Let $Q$ be the unique SE over $\singlewrgraph$. Then,
    \begin{enumerate}[label=(\roman*)]
        \item \label{itm:wr-wr0-1} if $Q$ is a singleton, the set of preferred strategies under the ND metric coincides with the set of all possible strategies returned by weakly better-response self-play, i.e., $\singlecomponentwr = \jointcomponentwr$;
        
        \item \label{itm:wr-wr0-2} if $Q$ is a non-singleton, the strategies returned by weakly better-response self-play must be preferred under the ND metric, i.e., $\jointcomponentwr  \subseteq \singlecomponentwr$. Moreover, there exists a two-player S-NFG $G$ such that there exists a preferred strategy which cannot be returned, i.e., $\jointcomponentwr  \subsetneq \singlecomponentwr$.
    \end{enumerate}
\end{thom}
\begin{proof}
    By Lemma \ref{lema:structure-wr0}, we have $\singlecomponentwr = Q$. Regarding \ref{itm:wr-wr0-1}, suppose that $Q = \{s_1\}$. Since $s_1$ is the unique SE, by Definition \ref{defi:ss-wbr} we have $J^1(s_1, s_3) > J^1(s_2, s_3)$ for all $s_2 \in \singlestrategyset\setminus \{s_1\}$ and $s_3 \in \singlestrategyset$ (if $\singlestrategyset = \{s_1\}$, then \ref{itm:wr-wr0-1} holds). By symmetry, we have $J^2(s_3, s_1) > J^2(s_3, s_2)$ for all $s_2 \in \singlestrategyset\setminus \{s_1\}$ and $s_3 \in \singlestrategyset$. We will use these payoff inequalities below.
    
    Next, we prove that there exists a (weakly better-response) path in $\jointwrgraph$ from any distinct joint strategy $(s_i, s_j) \in \jointstrategyset$ to $(s_1, s_1)$.  If $s_i = s_1$ and $s_j \neq s_1$, then $e_{(s_i, s_j)(s_1, s_1)} \in \jointwredge$ by $J^2(s_1, s_1) > J^2(s_i, s_j)$. If $s_i \neq s_1$ and $s_j = s_1$, we can prove it similarly. If $s_i \neq s_1$ and $s_j \neq s_1$, then $e_{(s_i, s_j)(s_1, s_j)} \in \jointwredge$ and $e_{(s_1, s_j)(s_1, s_1)} \in \jointwredge$ by $J^1(s_1, s_j) > J^1(s_i, s_j)$ and $J^2(s_1, s_1) > J^2(s_1, s_j)$, respectively. In conclusion, $(s_i, s_j)$ can reach $(s_1, s_1)$ along the edge $e_{(s_i, s_j)(s_1, s_1)}$ or the edges $(e_{(s_i, s_j)(s_1, s_j)}, e_{(s_1, s_j)(s_1, s_1)})$ in $\jointwrgraph$.
    
    Furthermore, since there is no outgoing edge for $(s_1, s_1)$ in $\jointwrgraph$, $(s_1, s_1)$ is the unique SE over $\jointwrgraph$, i.e, $\jointcomponentwr = \{s_1\}$. Therefore, we have $\jointcomponentwr  = \singlecomponentwr$.
    
    Regarding \ref{itm:wr-wr0-2},     
    let $\bar{\mathcal{G}}^{\textup{J}}_{\textup{WR}} = (\bar{{\jointstrategyset}}, \bar{\mathcal{E}}^{\textup{J}}_{\textup{WR}})$ be the subgraph of $\jointwrgraph$ such that $\bar{{\jointstrategyset}} = \{(s_1, s_2) \mid s_1 \in Q, s_2 \in Q \}$ and $e_{s_1^{\textup{J}} s_2^{\textup{J}}} \in \bar{\mathcal{E}}^{\textup{J}}_{\textup{WR}}$ if and only if $e_{s_1^{\textup{J}} s_2^{\textup{J}}} \in \jointwredge$ for all $s_1^{\textup{J}}, s_2^{\textup{J}} \in \bar{{\jointstrategyset}}$. Since $Q$ is the unique SE in $\singlewrgraph$, then by definition for $s_1 \in Q$ and $s_3 \in \singlestrategyset \setminus Q$, we have $J^1(s_1, s_4) > J^1(s_3, s_4)$ for all $s_4 \in \singlestrategyset$. By symmetry, we have $J^2(s_4, s_1) > J^2(s_4, s_3)$ for all $s_4 \in \singlestrategyset$. This implies that any walk over $\jointwrgraph$ cannot leave $\bar{\mathcal{G}}^{\textup{J}}_{\textup{WR}}$ once reaching it.

    Suppose, for the sake of contradiction, that there exists a strategy $s_1 \in \jointcomponentwr$ and $s_1 \notin \singlecomponentwr = Q$. Next, we prove that every joint strategy in $\jointstrategyset$ containing $s_1$ can reach $\bar{\mathcal{G}}^{\textup{J}}_{\textup{WR}}$ through the walks over $\jointwrgraph$ and cannot leave it forever.
 
    Given any strategy $s_2 \in \singlestrategyset$, we consider the joint strategy $(s_1, s_2)$, and the case $(s_2, s_1)$ follows similarly. If $s_2 \in Q$, we have $J^1(s_2, s_2) > J^1(s_1,s_2)$ noting that $Q$ is the unique SE over $\singlewrgraph$ and by assumption $s_1 \notin Q$. This implies that $e_{(s_1,s_2)(s_2, s_2)} \in \jointwredge$. Since $(s_2, s_2)$ occurs in $\bar{\mathcal{G}}_{\textup{WR}}^{\textup{J}}$, $(s_1, s_2)$ can reach $\bar{\mathcal{G}}_{\textup{WR}}^{\textup{J}}$ along the edge $e_{(s_1,s_2)(s_2, s_2)}$ over $\jointwrgraph$. On the other side, if $s_2 \notin Q$, then by taking another strategy $s_3 \in Q$, we have $e_{(s_1,s_2)(s_3, s_2)} \in \jointwredge$ and $e_{(s_3,s_2)(s_3, s_3)} \in \jointwredge$ using $J^1(s_3, s_2) > J^1(s_1,s_2)$ and $J^2(s_3, s_3) > J^2(s_3,s_2)$ respectively. Since $(s_3, s_3)$ occurs in $\bar{\mathcal{G}}_{\textup{WR}}^{\textup{J}}$, $(s_1, s_2)$ can reach $\bar{\mathcal{G}}_{\textup{WR}}^{\textup{J}}$ along the edges $(e_{(s_1,s_2)(s_3, s_2)}, e_{(s_3,s_2)(s_3, s_3)})$ over $\jointwrgraph$. In conclusion, $(s_1, s_2)$ is not included in any SE over $\jointwrgraph$. Noting that $s_2$ can be any strategy in $\singlestrategyset$, we obtain the contradiction that $s_1 \notin \jointcomponentwr$. Therefore, $\jointcomponentwr  \subseteq \singlecomponentwr$.
    
    Moreover, an example of two-player S-NFGs is provided in Fig. \ref{fig:SE-wr-wr0}, where $\singlecomponentwr = \{s_1, s_2\}$ and $\jointcomponentwr = \{s_1\}$. Therefore, we have $\jointcomponentwr \subsetneq \singlecomponentwr$.
\end{proof}

\subsection{Comparing Two Metrics}

Recall that the BD and ND metrics depend on different digraphs, and thus they might have different preferred strategies. Investigating the connections between their preferred strategies can shed light on the metric selection, and further on the selection of self-play variants in practice. 




\begin{thom}[Preferred strategies under the BD and ND metrics]\label{thm:se-relationship-gwr0-gt0}
    For a two-player S-NFG $G$, the preferred strategies under the BD metric are also preferred under the ND metric, i.e., $ \singlecomponentt \subseteq \singlecomponentwr$.
\end{thom}

\begin{figure}[t]
	\subfloat[]{    \label{fig:proof-of-se-relation-gwr0-gt0-a}%
	\begin{tikzpicture}
		\foreach \pos/\name in {{(0,0)/a},%
								{(-0.6,-1.04)/b},%
								{(0.6,-1.04)/c},%
								{(-0.6,-1.8)/d},%
								{(0.6,-1.8)/e},%
								{(-0.6, -3)/f},%
								{(0.6, -3)/g},%
								{(1.4,0)/h},%
								{(1.4,-1.04)/i},%
								{(1.4,-1.8)/j},%
								{(1.4,-3)/k}%
							}
        	\node[vertex] (\name) at \pos {};
        
        \draw[edge] (a) -- (b);
        \draw[edge] (b) -- (c);
        \draw[edge] (c) -- (a);
        \fill[color=green, opacity=0.2, rounded corners=7pt] (0,.4) -- (-1.0,-1.25) -- (1.0,-1.25) -- cycle;
        \node[anchor=west] () at (-1.65,-0.6) {$Q_1$};
        
        \draw[edge] (d) -- (f);
        \draw[edge] (f) -- (g);
        \draw[edge] (g) -- (e);
        \draw[edge] (e) -- (d);
        \fill[color=green, opacity=0.2, rounded corners=7pt] (-0.9, -1.55) -- (0.9, -1.55) -- (0.9, -3.3) -- (-0.9, -3.3) -- cycle;
        \node[anchor=west] () at (-1.65,-2.4) {$Q_2$};
        
        \draw[edge] (h) -- (a);
        \draw[edge] (i) -- (h);
        \draw[edge] (k) -- (g);
        \draw[edge] (j) -- (k);
        
	\end{tikzpicture}
}\hfil
\subfloat[]{    \label{fig:proof-of-se-relation-gwr0-gt0-b}%
	\begin{tikzpicture}
		\foreach \pos/\name in {{(0,0)/a},%
								{(-0.6,-1.04)/b},%
								{(0.6,-1.04)/c},%
								{(-0.6,-1.8)/d},%
								{(0.6,-1.8)/e},%
								{(-0.6, -3)/f},%
								{(0.6, -3)/g},%
								{(1.4,0)/h},%
								{(1.4,-1.04)/i},%
								{(1.4,-1.8)/j},%
								{(1.4,-3)/k}%
							}
        	\node[vertex] (\name) at \pos {};
        
        \draw[edge] (a) -- (b);
        \draw[edge] (b) -- (c);
        \draw[edge] (c) -- (a);
        \fill[color=green, opacity=0.2, rounded corners=7pt] (0,.4) -- (-1.0,-1.25) -- (1.0,-1.25) -- cycle;
        \node[anchor=west] () at (-1.65,-0.6) {$Q_1$};

        \draw[edge] (d) -- (f);
        \draw[edge] (f) -- (g);
        \draw[edge] (g) -- (e);
        \draw[edge] (e) -- (d);
        \draw[edge,MyRed,densely dashed] (d) -- (g);
        \fill[color=green, opacity=0.2, rounded corners=7pt] (-0.9, -1.55) -- (0.9, -1.55) -- (0.9, -3.3) -- (-0.9, -3.3) -- cycle;
        \node[anchor=west] () at (-1.65,-2.4) {$Q_2$};
        
        \draw[edge] (h) -- (a);
        \draw[edge] (i) -- (h);
        \draw[edge] (k) -- (g);
        \draw[edge] (j) -- (k);
      	\draw[edge,MyRed,densely dashed] (j) -- (i);
      	\draw[edge,MyRed,densely dashed] (i) -- (e);
        
	\end{tikzpicture}
}

\subfloat[]{    \label{fig:proof-of-se-relation-gwr0-gt0-c}%
	\begin{tikzpicture}
		\foreach \pos/\name in {{(0,0)/a},%
								{(-0.6,-1.04)/b},%
								{(0.6,-1.04)/c},%
								{(-0.6,-1.8)/d},%
								{(0.6,-1.8)/e},%
								{(-0.6, -3)/f},%
								{(0.6, -3)/g},%
								{(1.4,0)/h},%
								{(1.4,-1.04)/i},%
								{(1.4,-1.8)/j},%
								{(1.4,-3)/k}%
							}
      	\node[vertex] (\name) at \pos {};
        
        \draw[edge] (a) -- (b);
        \draw[edge] (b) -- (c);
        \draw[edge] (c) -- (a);
        
        \draw[edge] (d) -- (f);
        \draw[edge] (f) -- (g);
        \draw[edge] (g) -- (e);
        \draw[edge] (e) -- (d);
        \draw[edge,MyBlue] (d) -- (g);
        \draw[edge,MyRed,densely dashed] (b) -- (d);
        \draw[edge,MyRed,densely dashed] (e) -- (b);
        
        \fill[color=green, opacity=0.2, rounded corners=7pt] (0,.4) -- (-0.9,-1.05) -- (-0.9, -1.55) -- (-0.9, -3.3) -- (0.9, -3.3) -- (0.9, -1.55) -- (0.9,-1.05) -- cycle;
		\node[anchor=west] () at (-1.65,-1.6) {$Q$};
		
        \draw[edge] (h) -- (a);
        \draw[edge] (i) -- (h);
        \draw[edge] (k) -- (g);
        \draw[edge] (j) -- (k);
      	\draw[edge,MyBlue] (j) -- (i);
      	\draw[edge,MyBlue] (i) -- (e);
        
	\end{tikzpicture}
}\hfil
\subfloat[]{    \label{fig:proof-of-se-relation-gwr0-gt0-d}%
	\begin{tikzpicture}
		\foreach \pos/\name in {{(0,0)/a},%
								{(-0.6,-1.04)/b},%
								{(0.6,-1.04)/c},%
								{(-0.6,-1.8)/d},%
								{(0.6,-1.8)/e},%
								{(-0.6, -3)/f},%
								{(0.6, -3)/g},%
								{(1.4,0)/h},%
								{(1.4,-1.04)/i},%
								{(1.4,-1.8)/j},%
								{(1.4,-3)/k}%
							}
        	\node[vertex] (\name) at \pos {};
        
        \draw[edge] (a) -- (b);
        \draw[edge] (b) -- (c);
        \draw[edge] (c) -- (a);
        
        \draw[edge] (d) -- (f);
        \draw[edge] (f) -- (g);
        \draw[edge] (g) -- (e);
        \draw[edge] (e) -- (d);
        \draw[edge,MyBlue] (d) -- (g);
        \draw[edge,MyBlue] (b) -- (d);
        \draw[edge,MyBlue] (e) -- (b);
        \draw[edge,MyRed,densely dashed] (c) -- (i);
        
        \fill[color=green, opacity=0.2, rounded corners=7pt] (-0.1,0.3) -- (-0.9,-1.05) -- (-0.9, -1.55) -- (-0.9, -3.3) -- (0.9, -3.3) -- (0.9, -1.75) -- (1.65,-1.16) -- (1.65,0.3) -- cycle;
		\node[anchor=west] () at (-1.65,-1.6) {$Q'$};
        	        	        
        \draw[edge] (h) -- (a);
        \draw[edge] (i) -- (h);
        \draw[edge] (k) -- (g);
        \draw[edge] (j) -- (k);
      	\draw[edge,MyBlue] (j) -- (i);
      	\draw[edge,MyBlue] (i) -- (e);
  	\end{tikzpicture}
}
	\caption{An illustrative example for the proof of \thomref{thm:se-relationship-gwr0-gt0}. Newly added non-dominated edges are dashed in red, and all SEs are highlighted in green. 
    \subref{fig:proof-of-se-relation-gwr0-gt0-a} There are two SEs in the  best-response digraph $\singletgraph$. 
    \subref{fig:proof-of-se-relation-gwr0-gt0-b} Edges which do not change the SEs are added to the graph. 
    \subref{fig:proof-of-se-relation-gwr0-gt0-c} Edges between two SEs $Q_1$ and $Q_2$ are added, which merges these two SEs into a larger SE $Q$. 
    \subref{fig:proof-of-se-relation-gwr0-gt0-d} Edges from a node in $Q$ to a non-sink node enlarges the strategy set of all SEs, and finally the non-dominated digraph $\singlewrgraph$ is retained.}
    \label{fig:proof-of-se-relation-gwr0-gt0}
\end{figure}
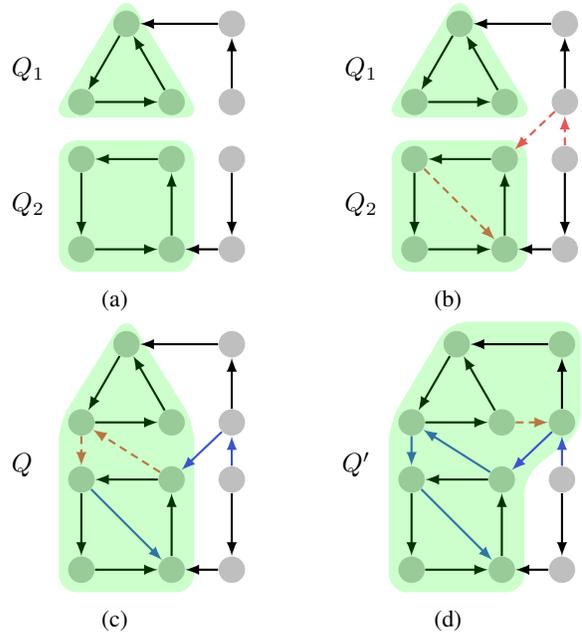

\begin{proof}
    By definition, $\singletgraph$ is a subgraph of $\singlewrgraph$. The theorem is proved by checking all SEs in $\singleallSEt$.
    
    Suppose that $|\singlewredge|-|\singletedge| = K\ge 0$, and by adding these $K$ non-dominated edges to $\singletgraph$, $\singlewrgraph$ is obtained. The order of adding these $K$ edges does not affect the SEs over $\singlewrgraph$, so the following order is used.
	
	To visualize the edge addition, an illustrative example is provided in Fig. \ref{fig:proof-of-se-relation-gwr0-gt0}, where the initial $\singletgraph$ has two SEs highlighted in green (Fig. \ref{fig:proof-of-se-relation-gwr0-gt0-a}). First, consider all non-dominated edges in $\singlewredge$ connecting a) two non-sink nodes in $\singletgraph$, b) two nodes in the same SE in $\singletgraph$, or c) a non-sink node to a sink node in $\singletgraph$, as indicated by red dashed edges in Fig. \ref{fig:proof-of-se-relation-gwr0-gt0-b}. Adding all these edges to $\singletedge$ has no impact on the SE, since they neither introduce a new SE nor remove any current one.
	
	Then, if $|\singleallSEt|>1$, we consider the edges in $\singlewredge$ between two distinct SEs $Q_1\in\singleallSEt$ and $Q_2\in\singleallSEt$, as Fig. \ref{fig:proof-of-se-relation-gwr0-gt0-c} shows. Take two strategies $s_1 \in Q_1$ and $s_2 \in Q_2$. Since $Q_1\in\singleallSEt$, there exists a strategy $s_1' \in Q_1$ ($s_1$ and $s_1'$ might be the same) such that $J^1(s_1, s_1') \ge J^1(s, s_1')$ for all $s \in \singlestrategyset$, implying that $J^1(s_1, s_1') \ge J^1(s_2, s_1')$. By definition, we have $e_{s_2s_1} \in \singlewredge$, i.e., there exists an edge in $\singlewredge$ from one node in $Q_2$ to one node in $Q_1$, as indicated by red dashed edges in Fig. \ref{fig:proof-of-se-relation-gwr0-gt0-c}. We have the similar conclusion from $Q_1$ to $Q_2$. Therefore, all SEs in $\singleallSEt$ are merged into a larger SE, denoted $Q$, through these edges in $\singlewredge$.

	Next, we consider edges in $\singlewredge$ from the merged SE $Q$ to a non-sink node $s$. Since $s$ is a non-sink node, there exists a path $(s, s_1, s_2, \ldots, s_k, s')$ $(k\in\mathbb{N}_{>0})$ from the non-sink node $s$ to $s'\in Q$. By adding an edge from $Q$ to $s$, the SE will be further enlarged as $Q' = Q\cup \{s, s_1, s_2, \ldots, s_k\}$, as in Fig. \ref{fig:proof-of-se-relation-gwr0-gt0-d}, and it is also true when adding an edge from $Q'$ to another non-sink node.
	
	With these edge addition, all non-dominated edges in $\singlewredge$ are added to $\singletedge$, which means that the digraph $\singlewrgraph$ is obtained. In this process the strategy set of all SEs monotonically expands. Thus, the strategy set of all SEs in $\singleallSEt$ is a subset of the strategy set in $\singleallSEwr$, which completes the proof.
\end{proof}

We provide an example to show that \thomref{thm:se-relationship-gwr0-gt0} has revealed the full connection between two metrics.

\begin{exmp}
    Consider the S-NFG in Fig. \ref{fig:SE-wr-wr0} again. By Fig. \ref{fig:brd-weakly-br-sp-self-single-graph}, we have $\singlecomponentwr = \{s_1, s_2\}$, and by drawing the best-response graph $\singletgraph$, we have $\singlecomponentt = \{s_1\}$. This implies that $ \singlecomponentt \subsetneq \singlecomponentwr$, i.e., there exists a preferred strategy of the ND metric which is not preferred by the BD metric.

\end{exmp}

Combining Theorems \ref{thm:se-relationship-gst-t0} and \ref{thm:se-relationship-gwr0-gt0}, we connect the ND metric and the strictly best-response self-play as follows.

\begin{cor}[ND metric and strictly best-response self-play]\label{corollary:se-relationship-gwr0-gst}
    If a two-player S-NFG $G$ has no mutual best-response pairs and no self best-response strategies, then all possible strategies returned by strictly best-response self-play are preferred under the ND metric, i.e., $ \jointcomponentst \subseteq \singlecomponentwr$.
\end{cor}


\subsection{Adjacency Matrices for $\singletgraph$ and $\jointstgraph$}\label{part:algebraic-relationship}


Our methods for strategy evaluation and learning crucially depend on the edges of the underlying digraphs. We next show that the adjacency matrices of $\singletgraph$ and $\jointstgraph$ are closely related through a class of digraph products \cite{west2001introduction}. Let $|\singlestrategyset| = n$ and $\singletadjmatrix=(a_{ij})_{n\times n}$ and $\jointstadjmatrix=(a_{ij})_{n^2\times n^2}$ be the adjacency matrices of $\singletgraph$ and $\jointstgraph$ respectively, where $a_{ij}=1$ if there is an edge from node $i$ to node $j$ in the corresponding digraph, and $a_{ij}=0$ otherwise.

Digraph products~\cite{west2001introduction} are commonly used to construct new families of possibly larger graphs from smaller ones. One of the most important and well-known digraph products is the \emph{digraph Cartesian product} as follows.


\begin{defi}[Digraph Cartesian product,~\cite{west2001introduction}]\label{def:digraph-cartesian-product}
    The Cartesian product of two digraphs $\mathcal{G}^1 = (\mathcal{S}^1,\mathcal{E}^1)$ and $\mathcal{G}^2 = (\mathcal{S}^2,\mathcal{E}^2)$, denoted $\mathcal{G}^1\square\mathcal{G}^2$, is a digraph $\jointstrategygraph =(\jointstrategyset,\jointstrategyedge)$ such that the set of nodes is $\jointstrategyset=\mathcal{S}^1\times\mathcal{S}^2$, and two nodes $s_1^{\textup{J}}=(s^1_1,s^2_1), s_2^{\textup{J}}=(s^1_2,s^2_2) \in \jointstrategyset$ are adjacent in $\jointstrategygraph$ (i.e., $e_{s_1^{\textup{J}}s_2^{\textup{J}}}\in\jointstrategyedge$), if and only if either 
    \begin{enumerate}
        \item $s^1_1=s^1_2$, and $e_{s^2_1s^2_2}\in\mathcal{E}^2$, or
        \item $s^2_1=s^2_2$, and $e_{s^1_1s^1_2}\in\mathcal{E}^1$.
    \end{enumerate}
\end{defi}

It is known \cite{west2001introduction} that if $\jointstrategygraph = \mathcal{G}^1\square\mathcal{G}^2$, then their adjacency matrices are related by the concise algebraic relation
\begin{equation}\label{eq:adjacency-matrix-digraph-cartesian-product}
    \bm{A}_{\jointstrategygraph} = \bm{I}_{n_1}\otimes\bm{A}_{\mathcal{G}^2} + \bm{A}_{\mathcal{G}^1}\otimes\bm{I}_{n_2},
\end{equation}
where $n_1 = |\mathcal{S}^1|$, $n_2 = |\mathcal{S}^2|$, and $\bm{A}_{\jointstrategygraph}$, $\bm{A}_{\mathcal{G}^1}$ and $\bm{A}_{\mathcal{G}^2}$ are the adjacency matrices of $\jointstrategygraph$, $\mathcal{G}^1$ and $\mathcal{G}^2$, respectively. For digraphs $\singletgraph=(\singlestrategyset,\singletedge)$ and $\jointstgraph=(\jointstrategyset,\jointstedge)$, the node sets satisfy the Cartesian product property, i.e., $\jointstrategyset = \singlestrategyset \times \singlestrategyset$. However, the following theorem shows that the edge conditions in Definition \ref{def:digraph-cartesian-product} are deviated by $\singletgraph$ and $\jointstgraph$, implying that $\jointstgraph \neq \singletgraph \square \singletgraph$. Noting this, we instead slightly modify the edge conditions in digraph Cartesian product such that $\singletadjmatrix$ and  $\jointstadjmatrix$ can be connected via a new relation similar to \eqref{eq:adjacency-matrix-digraph-cartesian-product}.




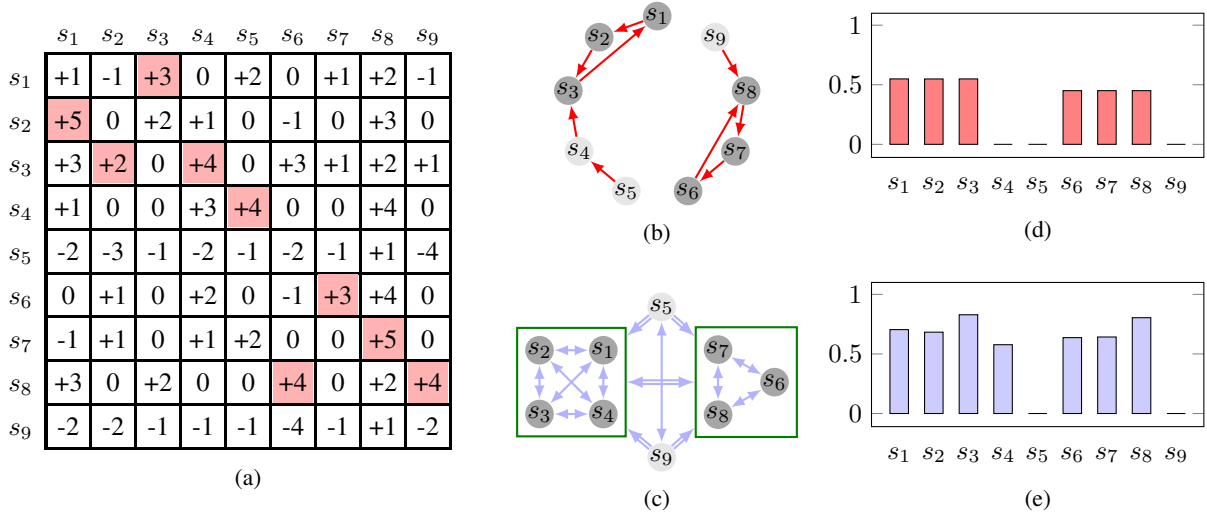
\begin{figure*}[t]
    \centering
    \centering
\begin{subcaptiongroup}
    \begin{minipage}[c]{.37\textwidth}
		\begin{game}{9}{9}[\captiontext*{}]\phantomcaption\label{fig:possible-result-alpha-rank-sink-based-metric-payoff-matrix}
				      \> $s_1$ \> $s_2$ \> $s_3$ \> $s_4$ \> $s_5$ \> $s_6$ \> $s_7$ \> $s_8$ \> $s_9$\\
				$s_1$\hspace{\fboxsep} \> +1 \> -1 \> \ts{+3} \> 0 \> +2 \> 0 \> +1 \> +2 \> -1 \\
				$s_2$\hspace{\fboxsep} \> \ts{+5} \> 0 \> +2 \> +1 \> 0 \> -1 \> 0 \> +3 \> 0 \\
				$s_3$\hspace{\fboxsep} \> +3 \> \ts{+2} \> 0 \> \ts{+4} \> 0 \> +3 \> +1 \> +2 \> +1 \\
				$s_4$\hspace{\fboxsep} \> +1 \> 0 \> 0 \> +3 \> \ts{+4} \> 0 \> 0 \> +4 \> 0 \\
				$s_5$\hspace{\fboxsep} \> -2 \> -3 \> -1 \> -2 \> -1 \> -2 \> -1 \> +1 \> -4 \\
				$s_6$\hspace{\fboxsep} \> 0 \> +1 \> 0 \> +2 \> 0 \> -1 \> \ts{+3} \> +4 \> 0 \\
				$s_7$\hspace{\fboxsep} \> -1 \> +1 \> 0 \> +1 \> +2 \> 0 \> 0 \> \ts{+5} \> 0 \\
				$s_8$\hspace{\fboxsep} \> +3 \> 0 \> +2 \> 0 \> 0 \> \ts{+4} \> 0 \> +2 \> \ts{+4} \\
				$s_9$\hspace{\fboxsep} \> -2 \> -2 \> -1 \> -1 \> -1 \> -4 \> -1 \> +1 \> -2
			\end{game}
	\end{minipage}\hfil
	\begin{minipage}[c]{.22\textwidth}
	    \centering
        \subcaptionbox{\label{fig:possible-result-alpha-rank-sink-based-metric-payoff-matrix-graphs-1}}{%
            \begin{tikzpicture}
    			\node[vertex,fill=black!35] (s1) at( 90 : 1.2cm) {$s_1$};
    			\node[vertex,fill=black!35] (s2) at(130 : 1.2cm) {$s_2$};
    			\node[vertex,fill=black!35] (s3) at(170 : 1.2cm) {$s_3$};
    			\node[vertex,fill=black!10] (s4) at(210 : 1.2cm) {$s_4$};
    			\node[vertex,fill=black!10] (s5) at(250 : 1.2cm) {$s_5$};
    			\node[vertex,fill=black!35] (s6) at(290 : 1.2cm) {$s_6$};
    			\node[vertex,fill=black!35] (s7) at(330 : 1.2cm) {$s_7$};
    			\node[vertex,fill=black!35] (s8) at( 10 : 1.2cm) {$s_8$};
    			\node[vertex,fill=black!10] (s9) at( 50 : 1.2cm) {$s_9$};
    			
    			\draw[edge,red] (s4) -- (s3);
    			
    			\draw[edge,red] (s1) -- (s2);
    			\draw[edge,red] (s2) -- (s3);
    			\draw[edge,red] (s3) -- (s1);
    			\draw[edge,red] (s5) -- (s4);

    			\draw[edge,red] (s9) -- (s8);
    			
    			\draw[edge,red] (s8) -- (s7);
    			\draw[edge,red] (s7) -- (s6);
    			\draw[edge,red] (s6) -- (s8);
    		\end{tikzpicture}
        }
        
        \vspace{1.71em}

        \subcaptionbox{\label{fig:possible-result-alpha-rank-sink-based-metric-payoff-matrix-graphs-2}}{%
            \begin{tikzpicture}
        		\node[vertex,fill=black!35,xshift=-1.2cm] (s1) at( 45 : 0.6cm) {$s_1$};
        		\node[vertex,fill=black!35,xshift=-1.2cm] (s2) at(135 : 0.6cm) {$s_2$};
        		\node[vertex,fill=black!35,xshift=-1.2cm] (s3) at(225 : 0.6cm) {$s_3$};
        		\node[vertex,fill=black!35,xshift=-1.2cm] (s4) at(315 : 0.6cm) {$s_4$};
        		
        		\draw[edge, blue!30, <->] (s1) -- (s2);
        		\draw[edge, blue!30, <->] (s1) -- (s3);
        		\draw[edge, blue!30, <->] (s1) -- (s4);
        		\draw[edge, blue!30, <->] (s2) -- (s3);
        		\draw[edge, blue!30, <->] (s2) -- (s4);
        		\draw[edge, blue!30, <->] (s3) -- (s4);
        
        		\node[condensed, fit=(s1) (s2) (s3) (s4), inner sep=1mm] (sinkp1) {};
        		
        		\node[vertex,fill=black!35,xshift=1.0cm] (s6) at( 0 : 0.5cm) {$s_6$};
        		\node[vertex,fill=black!35,xshift=1.0cm] (s7) at(120 : 0.5cm) {$s_7$};
        		\node[vertex,fill=black!35,xshift=1.0cm] (s8) at(240 : 0.5cm) {$s_8$};
        		
        		\draw[edge, blue!30, <->] (s6) -- (s7);
        		\draw[edge, blue!30, <->] (s7) -- (s8);
        		\draw[edge, blue!30, <->] (s6) -- (s8);
        		
        		\node[condensed, draw, fit=(s6)(s7)(s8), inner sep=1mm] (sinkp2) {};

        		\node[vertex, fill=black!10] (s5) at (0cm,1cm) {$s_5$};
        		\node[vertex, fill=black!10] (s9) at (0cm,-1cm) {$s_9$};
        
        		\draw[edge, blue!30, double, <->] (sinkp1) -- (sinkp2);
        		\draw[edge, double, blue!30] (s9) -- (sinkp1);
        		\draw[edge, double, blue!30] (s9) -- (sinkp2);
        		\draw[edge, double, blue!30] (s5) -- (sinkp1);
        		\draw[edge, double, blue!30] (s5) -- (sinkp2);
        		\draw[edge, blue!30, <->] (s9) -- (s5);  		
            \end{tikzpicture}
        }
	\end{minipage}\hfil
    \begin{minipage}[c]{.28\textwidth}
        \centering
        \subcaptionbox{\label{fig:possible-result-alpha-rank-sink-based-metric-payoff-matrix-result-1}}{%
            \begin{tikzpicture}[trim left]
		        \begin{axis}[
		        	height=3.5cm, width=6cm,
		        	ybar,
		        	bar width=7,
		        	ymax=1.1,
		        	symbolic x coords={$s_1$, $s_2$, $s_3$, $s_4$, $s_5$, $s_6$, $s_7$, $s_8$, $s_9$},
		        	xtick=data,
		        	major x tick style = transparent,
		        ]            
				\addplot[fill=red!50] coordinates {
					($s_1$, 0.5490)
					($s_2$, 0.5487)
					($s_3$, 0.5489)
					($s_4$, 0.0)
					($s_5$, 0.0)
					($s_6$, 0.4504)
					($s_7$, 0.4504)
					($s_8$, 0.4504)
					($s_9$, 0)
				}; 
		        \end{axis}
		  	\end{tikzpicture}	
        }
        
        \vspace{1em}
        
        \subcaptionbox{\label{fig:possible-result-alpha-rank-sink-based-metric-payoff-matrix-result-2}}{%
            \begin{tikzpicture}[trim left]
		        \begin{axis}[
		        	height=3.5cm, width=6cm,
		        	ybar,
		        	bar width=7,
		        	ymax=1.1,
		        	symbolic x coords={$s_1$, $s_2$, $s_3$, $s_4$, $s_5$, $s_6$, $s_7$, $s_8$, $s_9$},
		        	xtick=data,
		        	major x tick style = transparent,
		        ]
				\addplot[fill=blue!20] coordinates {
					($s_1$, 0.7041)
					($s_2$, 0.6827)
					($s_3$, 0.8287)
					($s_4$, 0.5774)
					($s_5$, 0.0)
					($s_6$, 0.6368)
					($s_7$, 0.6421)
					($s_8$, 0.8040)
					($s_9$, 0.0)
				}; 
		        \end{axis}
		  	\end{tikzpicture}
        }
    \end{minipage}
\end{subcaptiongroup}
    \caption{
    Strategy evaluation and learning for an abstract S-NFG $G$. \subref{fig:possible-result-alpha-rank-sink-based-metric-payoff-matrix} The payoff matrix is for the row player and its transpose is for the column player, where the maximum payoff of each column is highlighted in red. 
    \subref{fig:possible-result-alpha-rank-sink-based-metric-payoff-matrix-graphs-1} The best-response digraph $\singletgraph$ has two SEs which induce the preferred strategies of the BD metric. \subref{fig:possible-result-alpha-rank-sink-based-metric-payoff-matrix-graphs-2} The non-dominated digraph $\singlewrgraph$ has two SEs which induce the preferred strategies of the ND metric, where the self-loops are omitted for clarity and each double arrow indicates that the node has an arrow to all the nodes in the green rectangle.
    \subref{fig:possible-result-alpha-rank-sink-based-metric-payoff-matrix-result-1} The frequency of each strategy learnt by strictly best-response self-play.
    \subref{fig:possible-result-alpha-rank-sink-based-metric-payoff-matrix-result-2} The frequency of each strategy learnt by weakly better-response self-play.
    }
    \label{fig:possible-result-alpha-rank-sink-based-metric}
\end{figure*}

\begin{thom}[Adjacency matrices between $\singletgraph$ and $\jointstgraph$]\label{lemma:algebraic-relationship-gst-gt0}
    Consider the response digraphs $\singletgraph$ and $\jointstgraph$ of a two-player S-NFG $G$. Let $\singlestrategyset=\{1,2,\ldots,n\}$\footnote{For simplicity, this theorem adopts integers for strategies in $\singlestrategyset$.} and $\jointstrategyset=\{1,2,\ldots,n^2\}$ such that each node $r=(i-1)n+j \in\jointstrategyset$ consists of a strategy $i\in\singlestrategyset$ for player 1 and a strategy $j\in\singlestrategyset$ for player 2. Then, the adjacency matrices $\singletadjmatrix$ and $\jointstadjmatrix$ satisfy the condition
    \begin{equation}\label{eq:algebraic-relation}
        \jointstadjmatrix = \sum_{k=1}^n \left( \bm{e}_k\bm{e}_k^\top\otimes\adjmatrixbar_k + \adjmatrixbar_k\otimes\bm{e}_k\bm{e}_k^\top \right),
    \end{equation}
    where $\bm{e}_k \in \mathbb{R}^n$ and $\adjmatrixbar_k=\bm{1e}_k^\top\singletadjmatrix-\singletadjmatrix^\top\bm{e}_k\bm{e}_k^\top\singletadjmatrix$.
\end{thom}

\begin{proof}
    To avoid confusion, let $\jointstadjmatrix=(b_{rm})_{n^2\times n^2}$. Next, the theorem is proved by checking each element $b_{rm}$ ($r,m\in\jointstrategyset$) via Definitions \ref{def:strictly-best} and \ref{def:single-population-sst}, given \eqref{eq:algebraic-relation} holds. According to the indexes of nodes in $\singlestrategyset$ and $\jointstrategyset$, let $i_r,j_r,i_m,j_m\in\singlestrategyset$ such that $r=(i_r-1)n+j_r$ and $m=(i_m-1)n+j_m$. Thus for $b_{rm}$, 
\begin{equation}\label{eq:b-km}
\begin{aligned}
    &b_{rm}=(\bm{e}^\top_{i_r}\otimes\bm{e}^\top_{j_r}) \jointstadjmatrix (\bm{e}_{i_m}\otimes\bm{e}_{j_m})\\
    & = \sum_{k=1}^n (\bm{e}^\top_{i_r}\otimes\bm{e}^\top_{j_r})\big( \bm{e}_k\bm{e}_k^\top\otimes\adjmatrixbar_k + \adjmatrixbar_k\otimes\bm{e}_k\bm{e}_k^\top \big)(\bm{e}_{i_m}\otimes\bm{e}_{j_m})\\
    &=\sum_{k=1}^n \big(\bm{e}^\top_{i_r} \bm{e}_k\bm{e}_k^\top\bm{e}_{i_m}\otimes\bm{e}^\top_{j_r}\adjmatrixbar_k\bm{e}_{j_m} \\
    &\qquad \qquad +\bm{e}^\top_{i_r}\adjmatrixbar_k\bm{e}_{i_m}\otimes\bm{e}^\top_{j_r}\bm{e}_k\bm{e}_k^\top\bm{e}_{j_m} \big)\\
    &=\sum_{k=1}^n \big(\bm{e}^\top_{i_r} \bm{e}_k\bm{e}_k^\top\bm{e}_{i_m}\bm{e}^\top_{j_r}\adjmatrixbar_k\bm{e}_{j_m} +\bm{e}^\top_{i_r}\adjmatrixbar_k\bm{e}_{i_m}\bm{e}^\top_{j_r}\bm{e}_k\bm{e}_k^\top\bm{e}_{j_m} \big),
\end{aligned}
\end{equation}
where the first equality is by the index definition, the second equality follows from \eqref{eq:algebraic-relation}, the third equality follows from the mixed-product property of Kronecker product and the fourth equality is because the elements on both sides of $\otimes$ are $1$-by-$1$ dimensional. Next, several cases are discussed separately, depending on the relations among $i_r,j_r,i_m$ and $j_m$

Case 1: $i_r\neq i_m$ and $j_r\neq j_m$. For all $1\leq k\leq n$, we have
\begin{equation*}
    \bm{e}^\top_{i_r} \bm{e}_k\bm{e}_k^\top\bm{e}_{i_m}=0,\qquad \bm{e}^\top_{j_r}\bm{e}_k\bm{e}_k^\top\bm{e}_{j_m}=0,
\end{equation*}
and thus it follows from \eqref{eq:b-km} that $b_{rm}=0$, which satisfies~Definition \ref{def:strictly-best} because there is no edge from $r$ to $m$ in $\jointstgraph$ when two players have different strategies at $r$ and $m$.

Case 2: $i_r=i_m$ and $j_r\neq j_m$. By \eqref{eq:b-km}, $b_{rm}$ becomes
\begin{equation}\label{eq:b-km2}
\begin{aligned}
    b_{rm}&=\sum_{k=1}^n \Big(\bm{e}^\top_{i_r} \bm{e}_k\bm{e}_k^\top\bm{e}_{i_m}\bm{e}^\top_{j_r}\adjmatrixbar_k\bm{e}_{j_m}\Big)=\bm{e}^\top_{j_r}\adjmatrixbar_{i_r}\bm{e}_{j_m}\\
    &=\bm{e}^\top_{j_r}\big(\bm{1e}_{i_r}^\top\singletadjmatrix-\singletadjmatrix^\top\bm{e}_{i_r}\bm{e}_{i_r}^\top\singletadjmatrix\big)\bm{e}_{j_m}\\
    &=\bm{e}_{i_r}^\top\singletadjmatrix\bm{e}_{j_m}-\big(\bm{e}_{j_r}^\top\singletadjmatrix^\top\bm{e}_{i_r}\big)\big(\bm{e}_{i_r}^\top\singletadjmatrix\bm{e}_{j_m}\big)\\
    & =a_{i_rj_m}-a_{i_rj_r}a_{i_rj_m},
\end{aligned}
\end{equation}
where the first equality follows from $\bm{e}^\top_{j_r}\bm{e}_k\bm{e}_k^\top\bm{e}_{j_m}=0$ for all $1 \leq k \leq n$, the second equality is because $\bm{e}^\top_{i_r} \bm{e}_k\bm{e}_k^\top\bm{e}_{i_m}=1$ when $k = i_r$ and otherwise $\bm{e}^\top_{i_r} \bm{e}_k\bm{e}_k^\top\bm{e}_{i_m}=0$, the third equality is due to the definition of $\adjmatrixbar_{i_r}$ and the last equality follows from the index definition for $\singletadjmatrix$. If $a_{i_rj_m}=0$, then $b_{rm}=0$. This coincides with~Definition \ref{def:strictly-best}, because $a_{i_rj_m}=0$ implies that $j_m\notin\mathbb{B}(i_r)$, and thus there is no edge from $r$ to $m$ in  $\jointstgraph$. If $a_{i_rj_m}=1$ and $a_{i_rj_r}=0$, then $b_{rm}=1$. This is true because $a_{i_rj_m}=1$ and $a_{i_rj_r}=0$ imply that $j_m\in\mathbb{B}(i_r)$ and $j_r\notin\mathbb{B}(i_r)$, and thus there is an edge from $r$ to $m$ in  $\jointstgraph$.  If $a_{i_rj_m}=1$ and $a_{i_rj_r}=1$, then $b_{rm}=0$. This is true because $a_{i_rj_m}=1$ and $a_{i_rj_r}=1$ imply that $j_m\in\mathbb{B}(i_r)$ and $j_r\in\mathbb{B}(i_r)$, and thus there is no edge from $r$ to $m$ in  $\jointstgraph$. 

Case 3: $i_r=i_m$ and $j_r=j_m$. By \eqref{eq:b-km}, $b_{rm}$ becomes
\begin{equation*}
\begin{aligned}
    b_{rm}&=\bm{e}^\top_{j_r}\adjmatrixbar_{i_r}\bm{e}_{j_r}+\bm{e}^\top_{i_r}\adjmatrixbar_{j_r}\bm{e}_{i_r}\\
    & = \bm{e}^\top_{j_r} (\bm{1e}_{i_r}^\top\singletadjmatrix-\singletadjmatrix^\top\bm{e}_{i_r}\bm{e}_{i_r}^\top\singletadjmatrix) \bm{e}_{j_r} \\
    & \qquad \qquad \qquad \ \; + \bm{e}^\top_{i_r} (\bm{1e}_{j_r}^\top\singletadjmatrix-\singletadjmatrix^\top\bm{e}_{j_r}\bm{e}_{j_r}^\top\singletadjmatrix) \bm{e}_{i_r} \\
    &=\bm{e}_{i_r}^\top\singletadjmatrix\bm{e}_{j_r}-\bm{e}^\top_{j_r}\singletadjmatrix^\top\bm{e}_{i_r}\bm{e}_{i_r}^\top\singletadjmatrix\bm{e}_{j_r}\\
    &\qquad \qquad \qquad \ \; +\bm{e}_{j_r}^\top\bm{A}^{t0}\bm{e}_{i_r}-\bm{e}^\top_{i_r}\singletadjmatrix^\top\bm{e}_{j_r}\bm{e}_{j_r}^\top\singletadjmatrix\bm{e}_{i_r}\\
    & = a_{i_rj_r} - (\bm{e}^\top_{j_r}\singletadjmatrix^\top\bm{e}_{i_r}) (\bm{e}_{i_r}^\top\singletadjmatrix\bm{e}_{j_r})\\
    &\qquad \qquad \qquad \ \;  + a_{j_ri_r} - (\bm{e}^\top_{i_r}\singletadjmatrix^\top\bm{e}_{j_r}) (\bm{e}_{j_r}^\top\singletadjmatrix\bm{e}_{i_r}) \\
    &=a_{i_rj_r}-a_{i_rj_r}^2+a_{j_ri_r}-a_{j_ri_r}^2=0,
\end{aligned}
\end{equation*}
where the first equality is due to the second equality in \eqref{eq:b-km2}, and the last equality follows from the fact that $a_{i_rj_r},a_{j_ri_r}\in\{0,1\}$. This coincides with~Definition \ref{def:strictly-best}, because $i_r=i_m$ and $j_r=j_m$ implies that $r=m$, and there is no edge from $r$ to itself in $\jointstgraph$. The case $i_r\neq i_m$ and $j_r=j_m$ follows from the similar argument to Case 2.
\end{proof}

\begin{rek}
    The term $\singletadjmatrix^\top\bm{e}_k\bm{e}_k^\top\singletadjmatrix$ in $\adjmatrixbar_k$ is used to remove the self-loops which are not allowed in $\jointstgraph$ by~Definition \ref{def:strictly-best}. According to~\eqref{eq:adjacency-matrix-digraph-cartesian-product}, the Cartesian product of $\singletgraph$ (with itself) can be algebraically represented as 
    \begin{equation*}
    \begin{aligned}
        \bm{A}_{\singletgraph\square\singletgraph} & = \bm{I}_n\otimes\singletadjmatrix + \singletadjmatrix\otimes\bm{I}_n\\
        & = \sum_{k=1}^n \left( \bm{e}_k\bm{e}_k^\top\otimes\singletadjmatrix + \singletadjmatrix\otimes \bm{e}_k\bm{e}_k^\top \right).
    \end{aligned}
    \end{equation*}
    from which $\jointstadjmatrix \neq \bm{A}_{\singletgraph\square\singletgraph}$, i.e., $\jointstgraph \neq \singletgraph \square \singletgraph$.
\end{rek}




\section{Numerical Example}\label{part:outgrowths-of-the-main-results}

This section demonstrates the previous theoretical developments on a numerical example. In view of our focus on the strategy evaluation and learning, we consider an abstract two-player S-NFG instead of stochastic games directly.

Consider a two-player S-NFG $G$ with $9$ different strategies $\singlestrategyset = \{ s_i \}_{i=1}^9$ for each player, in which the row player's payoff matrix is given by Fig. \ref{fig:possible-result-alpha-rank-sink-based-metric-payoff-matrix} and the maximum payoff of each column is in red. We omit the column player's payoff matrix, as by symmetry, it is the transpose of the matrix in Fig. \ref{fig:possible-result-alpha-rank-sink-based-metric-payoff-matrix}.  The induced best-response digraph $\singletgraph$ and non-dominated digraph $\singlewrgraph$ are drawn in Figs. \ref{fig:possible-result-alpha-rank-sink-based-metric-payoff-matrix-graphs-1} and \ref{fig:possible-result-alpha-rank-sink-based-metric-payoff-matrix-graphs-2} respectively, where the self-loops are omitted for clarity and each double arrow in Fig. \ref{fig:possible-result-alpha-rank-sink-based-metric-payoff-matrix-graphs-2} indicates that the involved node has an arrow to all the nodes in the related green rectangle. Therefore by definition, the sets of preferred strategies under the BD and ND metrics, i.e., the strategies preferred for committing to the game system under the metrics, are
\[
\singlecomponentt = \{ s_1, s_2, s_3, s_6, s_7, s_8 \}, \singlecomponentwr = \{ s_1, s_2, s_3, s_4, s_6, s_7, s_8 \},
\]
respectively. Therefore, we have $\singlecomponentt \subsetneq \singlecomponentwr$.

With regard to the strategy learning, we consider the strictly best-response and weakly better-response self-plays both with maximum episode $\tau_\mathrm{max}=300$ and memory length $L=10$. The initial joint strategies are uniformly sampled from $\jointstrategyset = \singlestrategyset \times \singlestrategyset$. We run each variant of self-play for $10000$ times and then count how many times each strategy has been learnt in the final memory. The frequency of each strategy appearing in the final memory using the strictly best-response and weakly better-response self-plays is shown in Figs. \ref{fig:possible-result-alpha-rank-sink-based-metric-payoff-matrix-result-1} and \ref{fig:possible-result-alpha-rank-sink-based-metric-payoff-matrix-result-2} (bar charts), respectively. The statistics also align with the set of strategies occurring in the SEs over the joint strategy strictly best-response digraph $\jointstgraph$ and joint strategy weakly better-response digraph $\jointwrgraph$, respectively given by
\[
\begin{aligned}
\jointcomponentst & = \{ s_1, s_2, s_3, s_6, s_7, s_8 \}, \\
\jointcomponentwr & = \{ s_1, s_2, s_3, s_4, s_6, s_7, s_8 \}.
\end{aligned}
\]
Thus, $\jointcomponentst = \singlecomponentt$ and $\jointcomponentwr = \singlecomponentwr$, that is, the set of all possible strategies returned by strictly best-response and weakly better-response self-plays coincides with the set of preferred strategies under the BD and ND metrics, respectively.

\section{Conclusion}\label{part:conclusion}

We proposed the digraph-based BD and ND metrics for the strategy evaluation in two-player symmetric games which rank the strategies of a single player and thus help agents commit proper strategies to the game system. For the strategy learning, we introduced strictly best-response self-play which considers the strategy deviation with positive gain and weakly better-response self-play in which checking the conditions for changing strategies is efficient in practice. We proved that all possible learnt strategies by weakly better-response self-play are preferred under the ND metric, and all possible learnt strategies by strictly best-response self-play are preferred under the BD metric for a subclass of games. We demonstrated that the preferred strategies by the BD metric are also preferred by the ND metric. The adjacency matrix for the strictly best-response self-play is a new digraph product of that for the BD metric. Future works will involve strategy evaluation and learning before the commitment for multi-player asymmetric games.



\bibliographystyle{IEEEtran}
\bibliography{ref}

\end{document}